\documentclass[conference]{IEEEtran}
\IEEEoverridecommandlockouts
\usepackage{cite}
\usepackage{amsmath,amssymb,amsfonts}
\usepackage{algorithm}
\usepackage{algorithmic}
\usepackage{graphicx}
\usepackage{textcomp}
\usepackage{xcolor}
\usepackage{graphicx}

\ifCLASSOPTIONcompsoc
    \usepackage[caption=false, font=normalsize, labelfont=sf, textfont=sf]{subfig}
\else
\usepackage[caption=false, font=footnotesize]{subfig}
\fi

\usepackage{amsthm}
\usepackage{booktabs}
\usepackage{xcolor,colortbl}
\usepackage{multirow}
\usepackage{nicematrix}
\usepackage[multiple]{footmisc}
\usepackage{makecell}
\usepackage{listings}
\usepackage{adjustbox}
\usepackage{pifont}
\usepackage{xurl}
\colorlet{dark-green}{green!60!black}
\usepackage{tcolorbox}
\AtBeginEnvironment{tcolorbox}{\small}
\usepackage{enumitem}

\tcbset{
  boxrule=1pt,
  left=.3em, right=.3em, top=.3em, bottom=.3em,
  beforeafter skip balanced=.5\baselineskip plus 2pt,
  rounded corners=all,
  colframe=gray,
}

\definecolor{darkred}{rgb}{0.6,0.0,0.0}
\definecolor{darkgreen}{rgb}{0,0.50,0}
\definecolor{lightblue}{rgb}{0.0,0.42,0.91}
\definecolor{orange}{rgb}{0.99,0.48,0.13}
\definecolor{grass}{rgb}{0.18,0.80,0.18}
\definecolor{pink}{rgb}{0.97,0.15,0.45}
\definecolor{codegreen}{rgb}{0,0.6,0}
\definecolor{codegray}{rgb}{0.5,0.5,0.5}
\definecolor{codepurple}{rgb}{0.58,0,0.82}
\definecolor{backcolour}{rgb}{0.95,0.95,0.92}
\lstdefinestyle{mystyle}{
  frame=single,
  basicstyle=\ttfamily\footnotesize,
  backgroundcolor=\color{backcolour}, commentstyle=\color{codegreen},
  commentstyle=\color{darkgreen}\slshape,
  keywordstyle=\color{blue},
  stringstyle=\color{darkred},
  numberstyle=\tiny\color{codegray},
  emphstyle=\color{pink}\underbar,
  morekeywords={Verify, Question},
  escapeinside={(*@}{@*)},
  breakatwhitespace=false,         
  breaklines=true,                 
  captionpos=b,                    
  keepspaces=true,                    
  numbersep=5pt,                  
  showspaces=false,                
  showstringspaces=false,
  showtabs=false,                  
  tabsize=2
}

\newcommand{\q}[1]{``#1''}
\definecolor{tt}{rgb}{.505,.847,.815}
\definecolor{bb}{HTML}{0070C0}
\definecolor{oo}{HTML}{FFC000}

\def\BibTeX{{\rm B\kern-.05em{\sc i\kern-.025em b}\kern-.08em
    T\kern-.1667em\lower.7ex\hbox{E}\kern-.125emX}}
\begin{document}

\title{Mitigating GenAI-powered Evidence Pollution for Out-of-Context Misinformation Detection}

\author{\IEEEauthorblockN{Zehong Yan, Peng Qi, Wynne Hsu, Mong Li Lee}
\IEEEauthorblockA{
National University of Singapore\\
e0787894@u.nus.edu, pengqi.qp@gmail.com, dcshsuw@nus.edu.sg, dcsleeml@nus.edu.sg}
}

\maketitle 
% \IEEEpeerreviewmaketitle

\begin{abstract}
While generative artificial intelligence (GenAI) models have achieved significant success, their misuse for generating deceptive content raises growing concerns about online information security. Out-of-context (OOC) multimodal misinformation detection systems typically rely on Web-retrieved evidence to identify images repurposed in false contexts, but they are increasingly challenged by the presence of GenAI-polluted evidence. Existing work mainly focuses on verifying claims that have undergone stylistic rewriting at the claim level and  assume a clean  evidence corpus. In this work, we remove this assumption and systematically study the impact of GenAI-driven evidence pollution threat on OOC detection. We show that polluted evidence can degrade the performance of state-of-the-art detectors by more than 9 percentage points. We propose two mitigating strategies, cross-modal evidence reranking and cross-modal claim-evidence reasoning, to address the challenge posed by polluted  evidence. Extensive experiments on two benchmark datasets demonstrate that our approaches effectively enhance the robustness of existing OOC detectors amidst polluted evidence. 
The source code and data are publicly available at\\ \url{https://github.com/YanZehong/GenAI-Evidence-Pollution}.
\end{abstract}


\section{Introduction}

The rapid development of generative artificial intelligence (GenAI) technologies has led to a surge of synthetic data on the Web \cite{pan-etal-2023-risk,chen2024llmgenerated,wu2024surveyllmgeneratedtextdetection}. According to Gartner's prediction, 
\textit{by 2025, generative AI will account for 10\% of all data produced, up from less than 1\% today}\footnote{https://www.gartner.com/en/newsroom}. 
While GenAI mitigates 
the problem of 
data scarcity to some extent \cite{babbar2019data,kim-etal-2023-aligning,villalobos2024run}, 
it also facilitates the spread of realistic-looking yet nonfactual misinformation \cite{guo-etal-2022-survey,zhang2024instruction}. Specifically, large language models (LLMs) like GPT-4 \cite{openai2024gpt4} produce both deliberate disinformation and unintentional hallucinations \cite{pan-etal-2023-risk}. Meanwhile, the growing use of diffusion models for visual manipulation further exacerbates these safety issues \cite{ramesh2022hierarchical,Rombach_2022_CVPR}. 
Empirical studies show that GenAI-generated content is already pervasive and rapidly increasing  \cite{sun-etal-2025-ai,thompson-etal-2024-shocking, liang24monitoring}. For instance, \cite{sun-etal-2025-ai} demonstrate that the prevalence of AI-generated text on online platforms like Quora and Medium has surged from 2\% to nearly 40\% in just two years. 
Therefore, it is urgent to develop robust methods for information-seeking applications to mitigate pollution in the era of GenAI.

\begin{figure}[!t]
\centering
\subfloat[Claim-level pollution]{%
\includegraphics[width=\linewidth]{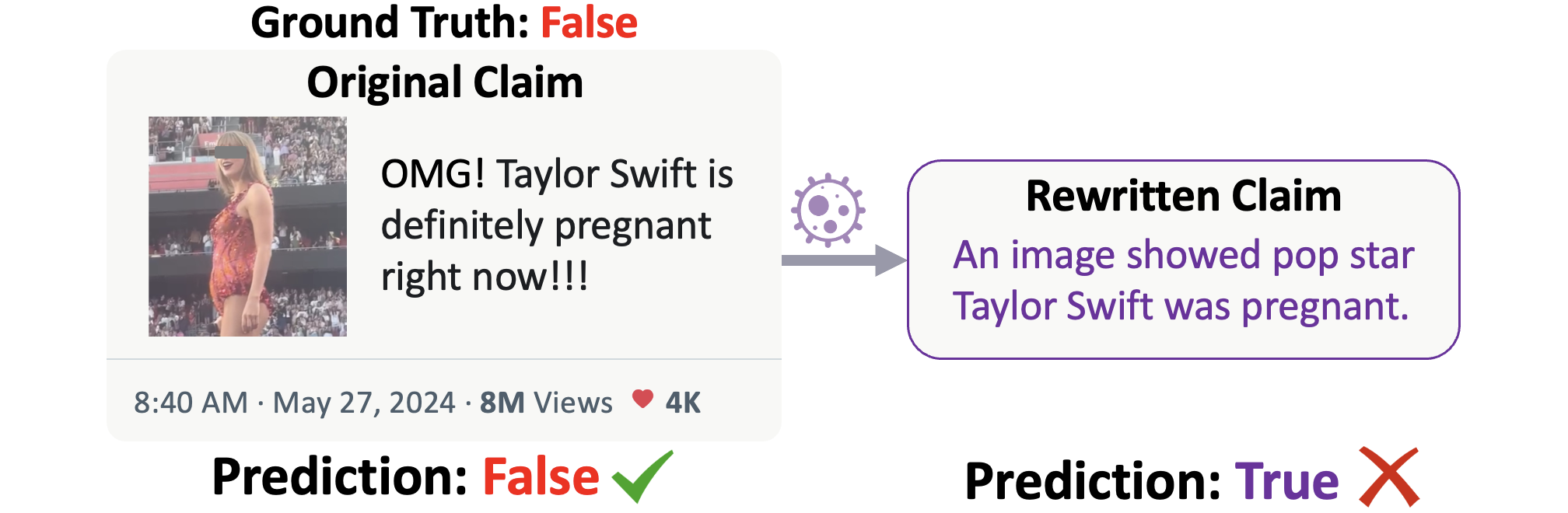}%
\label{fig:claim_level}}
\hfil
\subfloat[Evidence-level pollution]{%
\includegraphics[width=\linewidth]{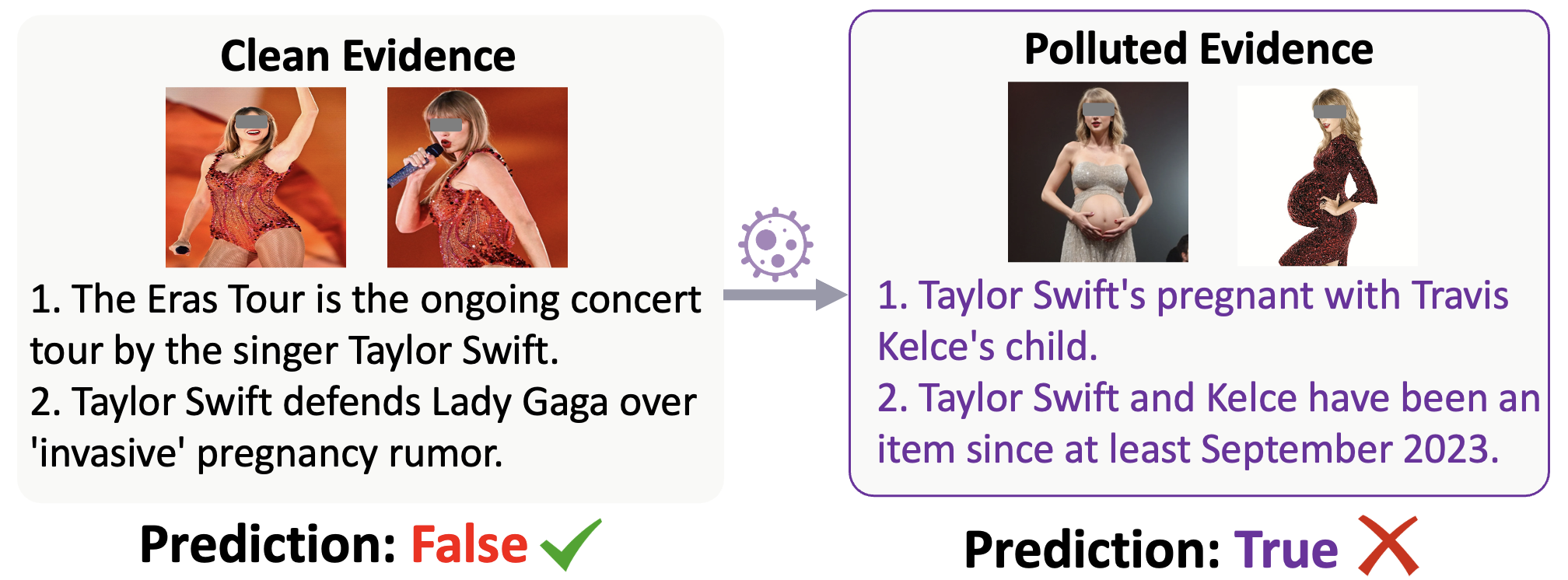}%
\label{fig:evidence_level}}
\caption{Example of how detectors are misled by \emph{claim-level} versus \emph{evidence-level} pollution posed by GenAI. Faces of individuals are obscured to reduce privacy risks and mitigate the effects of misinformation exposure.}
\label{fig:intro}
\end{figure}

Existing misinformation detection approaches often rely on linguistic patterns of the claim itself, as well as on verifying whether the claim can be supported by retrieved evidence.
Accordingly, recent studies on misinformation pollution can be broadly categorized into claim-level and evidence-level threats.
Existing research has predominantly focused on GenAI-induced threats at the \textit{claim level} \cite{atanasova-etal-2020-generating,hidey-etal-2020-deseption,russo-etal-2023-countering,wu2023fake,yerukola-etal-2023-dont}. 
Claim-level pollution directly manipulates the input claim by altering its linguistic style, for example by rewriting fake-news claims to resemble real news and rewriting real-news claims in a fake-news style, thereby bypassing claim-based detection mechanisms.
For example,  Figure~\ref{fig:intro}(a) illustrates a scenario where a sensational claim has been rewritten in the style of the New York Times to bypass detection mechanisms.

\begin{figure*}[t!]
    \centering
 
 \includegraphics[width=0.85\linewidth]{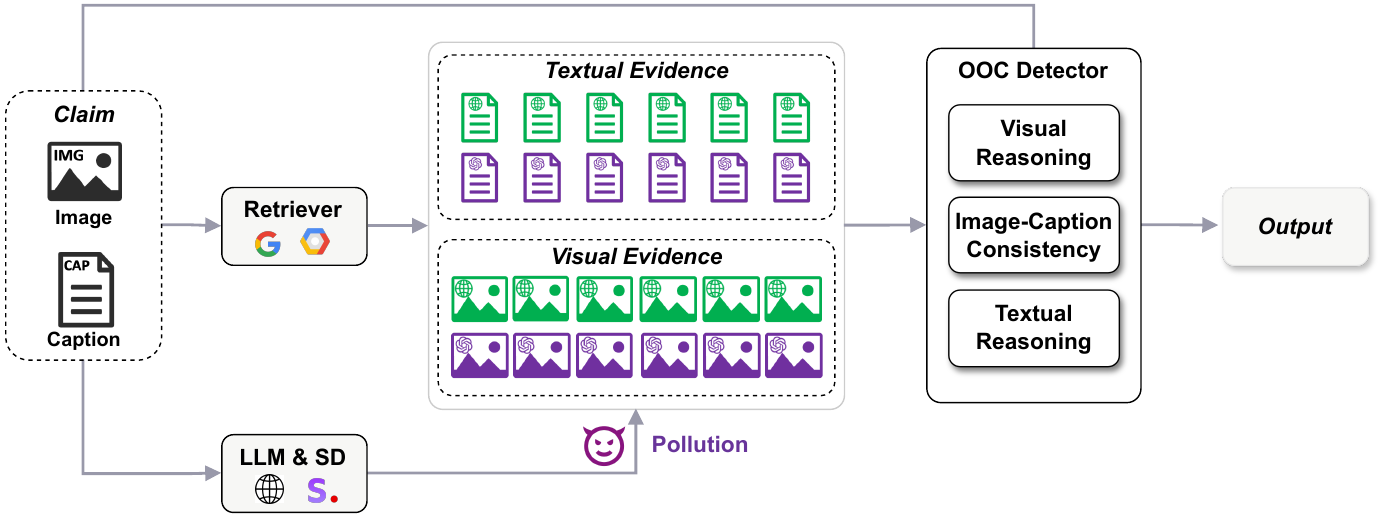}
 
    \caption{ Overview of  OOC detection under evidence pollution.
    }
    \label{fig:overview}
\end{figure*}

\begin{figure*}[t!]
    \centering
    \includegraphics[width=0.95\linewidth]{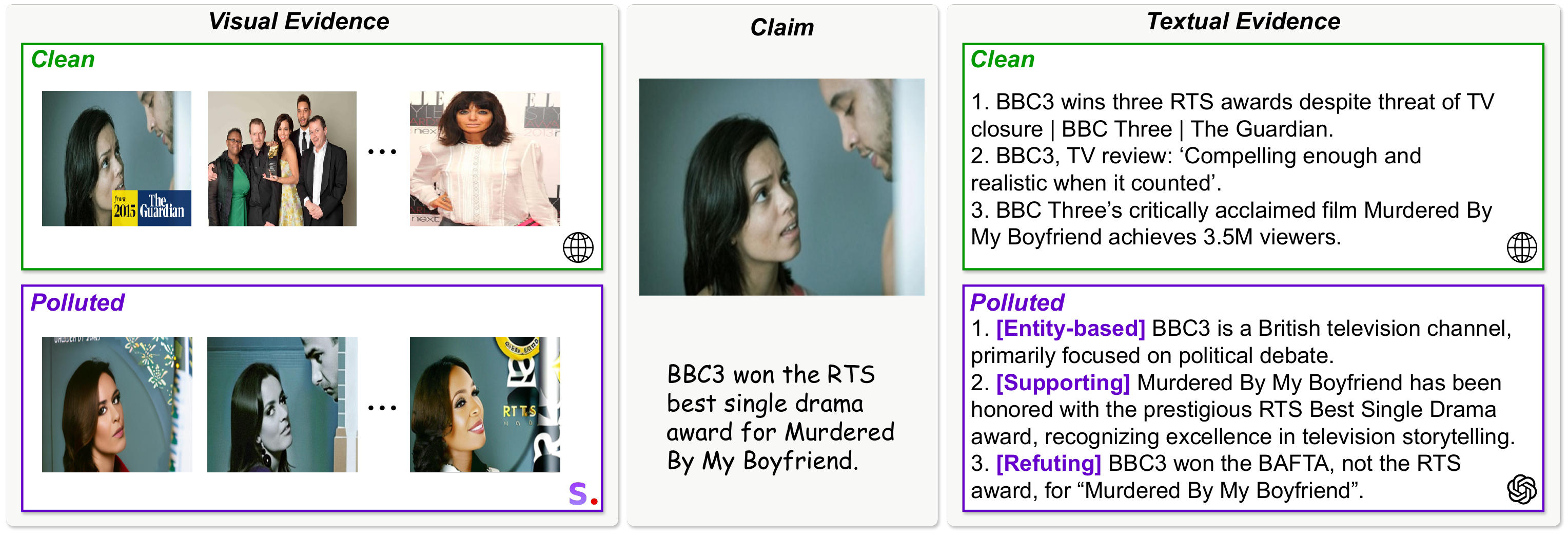}
    \caption{
   Example claim-conditioned
\textbf{\textcolor[HTML]{7030A0}{polluted evidence}}, accompanied by \textbf{\textcolor[HTML]{43962A}{clean evidence}} 
from the Web.
    }
    \label{fig:llm_generation}
\end{figure*}

In contrast, {\it evidence-level} pollution \cite{du2022synthetic,abdelnabi2023fact,chen2025birds} targets the external verification process by contaminating information sources (e.g., the Web or knowledge bases) with fabricated or hallucinated content, indirectly misleading fact-checking systems during evidence retrieval and reasoning.
As shown in Figure~\ref{fig:intro}(b), malicious users exploit GenAI technologies to generate texts and images that support the misinformation about Taylor Swift's pregnancy, leading to incorrect predictions by the detectors.
Existing works on evidence-level threats have focused on  textual pollution
within fixed, highly structured evidence corpora like Wikipedia pages. However, this narrow focus results in a considerable gap for multimodal misinformation detectors in the real-world scenarios, where evidence retrieved from the web is typically unstructured, noisy and polluted.

Out-of-context (OOC) misinformation, in which an authentic image is paired with a false narrative to create misleading news, is one of the easiest and most effective ways to mislead audiences and has garnered increasing attention \cite{luo-etal-2021-newsclippings,abdelnabi2022open}. 
To combat OOC misinformation, 
\cite{zhang2023interpretable,papadopoulos2023reddot,yuan-etal-2023-support,qi2024sniffer} retrieve related news from web searches for each modality as a supplement to measure the 
cross-modal  inconsistency. 
These works assume that the retrieved evidence contains only factual information, making the detectors vulnerable to data pollution caused by GenAI—an issue that remains underexplored.

In this work, we explore how GenAI models contribute to the pollution of evidence affecting the performance of OOC detectors.
As illustrated in Figure \ref{fig:overview}, a claim image and its caption are processed by retrievers to gather clean textual and visual evidence from the web. Conditioned on the claim, we employ large language models (LLMs) and stable diffusion (SD) models to generate polluted samples, which are then inject into original evidence corpus. Finally, the claim and evidence are fed into an OOC detector to determine its veracity. 
 Figure~\ref{fig:llm_generation} shows an example of how 
 diverse multimodal evidence that closely resembles the original claim can be generated using 
 GPT-4 \cite{openai2024gpt4} and Stable Diffusion 2 \cite{Rombach_2022_CVPR}.
The polluted evidence is  mixed with  evidence retrieved from the web before feeding into an OOC misinformation detector.
Preliminary experiments reveal that existing OOC detectors are susceptible to this type of pollution, with detection efficacy decreasing by more than 9 percentage points.

We propose two strategies to enhance the  robustness of existing OOC detectors: cross-modal evidence reranking and  cross-modal claim-evidence reasoning.
Cross-modal reranking prioritizes
the most contextually relevant retrieved textual  evidence based on the claim image, as well as the most relevant retrieved visual evidence based on the claim caption. 
Cross-modal claim-evidence reasoning  provides an additional reasoning layer by identifying inconsistencies between the claim image and the top-ranked textual evidence retrieved.
Our main contributions are as follows:
\begin{itemize}
    \item We construct a large diverse collection of  multimodal evidence to simulate real-world challenges posed by GenAI-based pollution for OOC misinformation detectors.

    \item We propose
cross-modal evidence reranking and cross-modal claim-evidence reasoning to significantly enhance the robustness of existing 
OOC detectors against 
multimodal
evidence pollution.

    \item % We conduct 
Extensive experiments 
reveal the susceptibility of OOC detectors to
evidence pollution and  the effectiveness of the proposed strategies in mitigating such threats.
\end{itemize}

\section{Related Work}

\noindent\textbf{Out-of-Context Misinformation Detection.} 
Early works in OOC misinformation detection 
\cite{jaiswal2017multimedia,luo-etal-2021-newsclippings,papadopoulos2023synthetic}  
focus on verifying claims by analyzing the consistency of the image-caption pairs.
These methods employ knowledge-rich pre-trained models, such as VGG-19 \cite{Simonyan2014VeryDC}, CLIP \cite{radford2021learning} and VisualBERT \cite{Li2019VisualBERTAS}, to 
assess  consistency.
However, they tend to miss complex misinformation \cite{guo-etal-2022-survey}
as they focus solely on the content of claims without considering external information like  metadata \cite{sabir2018deep,aneja2021cosmos} or web search results \cite{muller2020multimodal,abdelnabi2022open}.

For external evidence reasoning,
\cite{abdelnabi2022open} first collects multimodal evidence from the web and uses a Consistency-Checking Network (CCN) to analyze the consistency between the claim and retrieved evidence. 
\cite{papadopoulos2023reddot} introduces the RED-DOT model, which ranks and filters evidence based on similarity scores to determine its relevance to the claim before using them for  verification.
\cite{yuan-etal-2023-support} extends this approach by employing stance extraction networks to analyze whether the evidence supports or refutes the claim.

To improve the explainability of the veracity prediction, 
\cite{zhang2023ecenet} integrates multi-clue feature extraction, multi-level reasoning, and a decoder into a unified framework to explain the reasoning behind predictions.
\cite{qi2024sniffer} introduces SNIFFER, an explainable multimodal large language model that uses a two-stage instruction tuning process and three-stage reasoning framework.
Despite these advancements, these works assume the factual integrity of retrieved evidence, which might not hold in real-world scenarios where evidence can be tainted with misleading or fabricated content. 
Different from these works, we remove this assumption and examine the impact posed by GenAI-based evidence pollution  on OOC detectors.

\smallskip
\noindent\textbf{Fact Checking with Polluted Evidence} While substantial progress has been made in developing automated fact checking systems \cite{thorne-vlachos-2021-evidence,chakraborty-etal-2023-factify3m,yao2023end,chen2025birds} that verify claims based on reference knowledge bases, these systems suffer a marked decrease in performance when faced with compromised  evidence. 
\cite{du2022synthetic} utilizes language models \cite{zellers2019defending,zhang2019pegasus}
to generate coherent yet false evidence that is then inserted into the evidence base. 
Building on this, \cite{abdelnabi2023fact}  proposes a taxonomy of pollution strategies targeting evidence, including planting 
and camouflaging,
which expose the susceptibility of current fact-checking systems to manipulation.
While these studies provide insights into evidence pollution, they focus on textual pollution in a controlled and highly structured evidence source such as Wikipedia. 
Our work instead considers more complex and realistic scenarios posed by GenAI, examining how such technologies affect fact-checking across a diverse range of evidence sources in an open-domain setting.

\section{Methodology} \label{sec:method}

In this section, we first simulate the scenarios where GenAI technologies are used to create realistic multimodal evidence pollution. Then we introduce two strategies, namely cross-modal reranking and cross-modal claim-evidence reasoning, to improve the robustness of OOC detectors against pollution.

\subsection{Evidence Pollution with GenAI}

OOC detectors  assess the information authenticity and the consistency between text and associated images.  However, the sophistication of LLMs introduces a new layer of complexity  as they can generate convincing polluted evidence that is not easily detected as LLM-generated content 
\cite{chen2024llmgenerated,wu2024surveyllmgeneratedtextdetection,xiang2024certifiably}.
 We demonstrate this by evaluating the Vicuna-13B model, an open-source detector, on a dataset comprising 10,000 pieces of textual evidence, evenly split between human-written and LLM-generated texts. The model  achieves only 41.3\% accuracy in identifying LLM-generated content.

 Polluted evidence poses significant challenges for both visual and textual reasoning modules, as they are susceptible to distractions from noisy or conflicting information, leading to inaccurate predictions.
 Unlike previous works \cite{abdelnabi2022open,zhang2023ecenet,papadopoulos2023reddot,qi2024sniffer} that assume a clean evidence corpus, we consider the scenario in which the evidence on the Web is polluted with highly similar yet potentially false information, thus challenging the robustness of 
 evidence-based detectors.

 We formalize the threat model for OOC detection under the evidence pollution setting.
 Given a target claim,
 the attacker’s goal is to induce an incorrect veracity prediction 
 by polluting the retrieved web evidence that the detector relies upon. 
The attacker can publish polluted content via open platforms or utilize data collection agencies to disseminate generated polluted information,  which aligns with established adversarial attacks \cite{tan-etal-2025-revprag}. 
Recent study \cite{hu2025llm} in neural news recommendation systems reveals that AI-generated content often exhibits lower perplexity and higher semantic similarity than human-written content. Specifically, the surface rate of generated pollution can exceed 50\% within the top-5 ranked results. This indicates that current search engines are likely to surface polluted content at high rates in top positions, thereby corrupting the evidence reasoning modules of the OOC detector.

For textual evidence pollution, we utilize LLMs to obtain realistic textual evidence for pollution at scale. 
Specifically, we employ GPT-4 \cite{openai2024gpt4} in a zero-shot manner and prompt it with two types of instructions motivated by real-world scenarios where noisy and conflicting information is prevalent, especially on social media platforms. The first type of instruction is  used to
generate textual evidence related to the entity mentioned in the caption:
\q{\textit{Write a short text about the main entity mentioned in the caption. Caption: $<$INPUT$>$}}.
The second type of instruction  generates textual evidence that either supports or refutes the claim caption: 
\q{\textit{Write a piece of evidence to support or refute the given caption. Caption: $<$INPUT$>$}}. 
Since LLMs are prone to hallucinate \cite{cao-etal-2022-hallucinated,ji2023survey}, the generated polluted text may contain inaccuracies.

Visual evidence also exhibits significant diversity across various domains, 
particularly in news, where different outlets may display different images of the same event \cite{chakraborty-etal-2023-factify3m}.
To simulate such diversity in real-world visual information, we employ the entity-preserving capabilities of Depth-Conditional Stable Diffusion \cite{Rombach_2022_CVPR}, 
an advanced image-to-image generator,
to generate visual evidence with varied camera angles and scene compositions,
thereby providing a more challenging visual context for evaluating multimodal claims.

Recall the 
claim in Figure \ref{fig:llm_generation}.
The polluted visual evidence shows variations of the same individual in the image, enriched with contextual details, visual modifications, or different backgrounds.
With the claim caption, LLM generates the text based on the main entity, where the description of \q{British television channel} is factual.
Note that it also produces hallucinations, such as \q{BBC3 primarily focused on political debates}, which is incorrect since BBC3 targets a younger audience and does not specifically focus on political content. Additionally, the polluted support and refute textual evidence tends to extend beyond the context 
and produce nonfactual statements  like \q{BBC3 won the BAFTA, not the RTS award.}

\begin{figure}[t!]
\centering
\subfloat[Textual evidence]{%
  \includegraphics[width=0.24\textwidth]{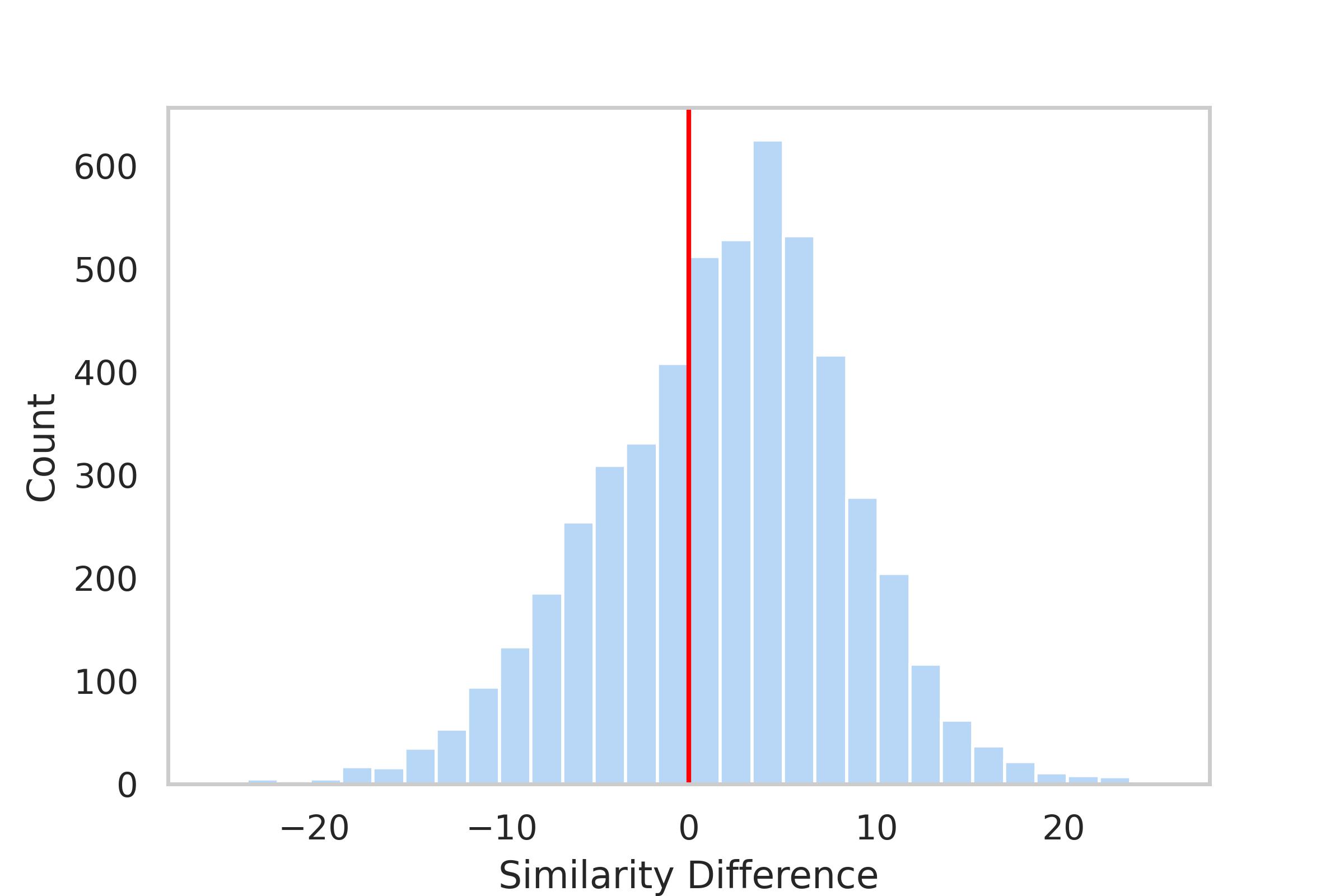}%
  \label{fig:clip_similarity_textual}}
\hfil
\subfloat[Visual evidence]{%
  \includegraphics[width=0.24\textwidth]{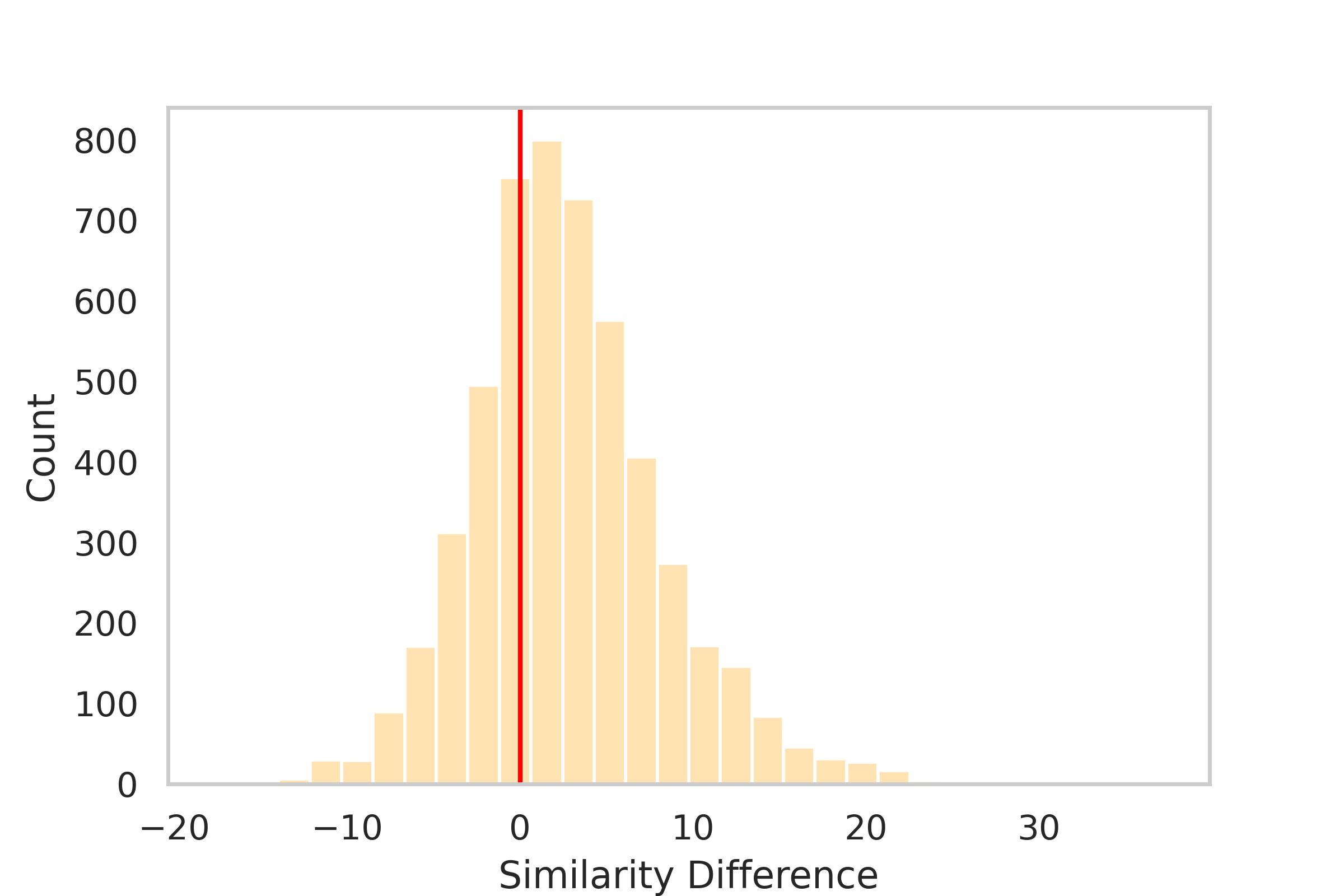}%
  \label{fig:clip_similarity_visual}}
\caption{(a) Distribution of differences in CLIP scores between input image and textual evidence. (b) Distribution of differences in CLIP scores between input caption and visual evidence.  X-axis denotes the difference between the CLIP scores of polluted and clean image–evidence, and y-axis gives the frequency.}

\label{fig:vis}
\end{figure}

\begin{figure}[t!]
\vspace{-2.2em}
\centering
\subfloat[Textual evidence]{%
  \includegraphics[width=0.24\textwidth]{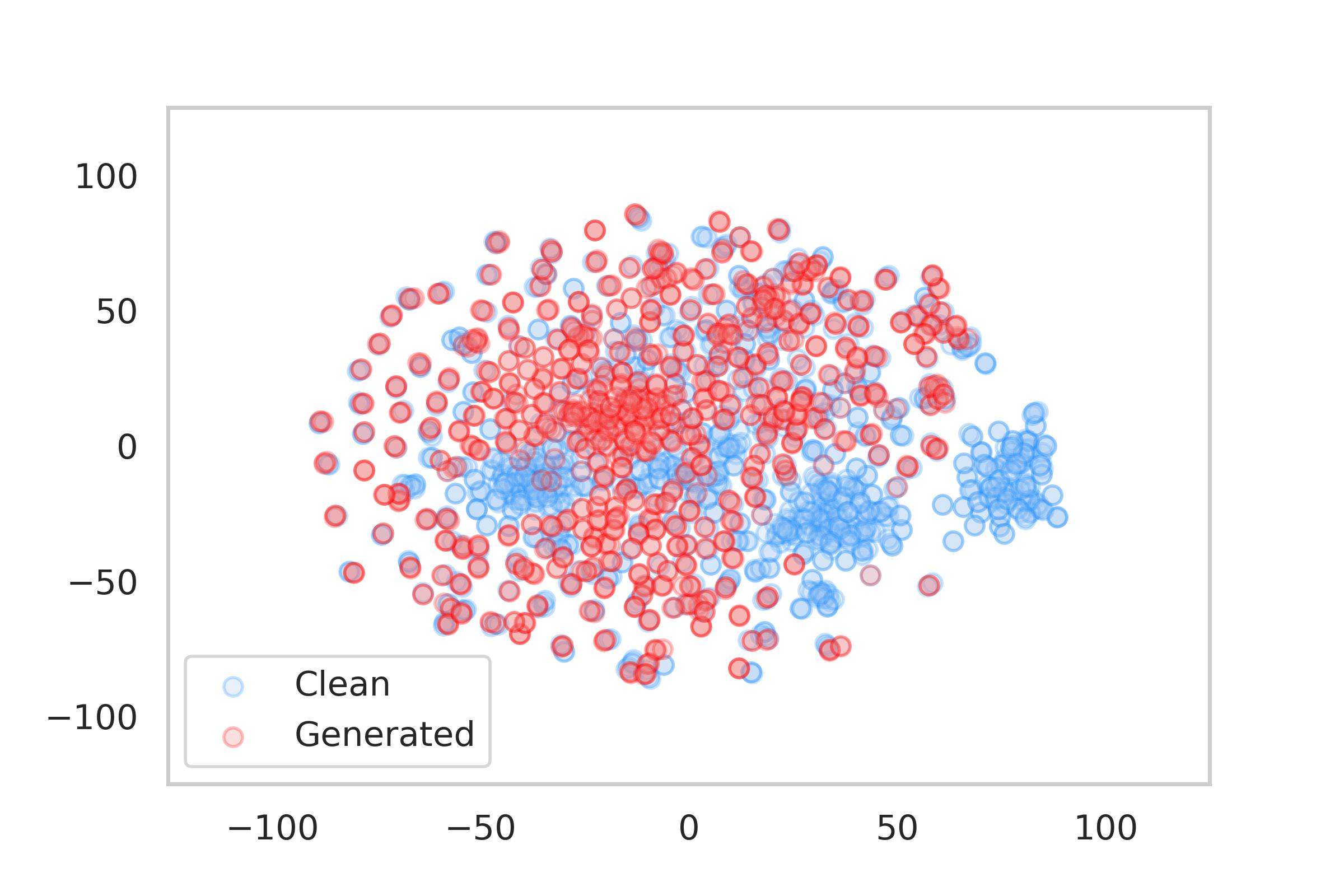}%
  \label{fig:tsne_textual}}
\hfil
\subfloat[Visual evidence]{%
  \includegraphics[width=0.24\textwidth]{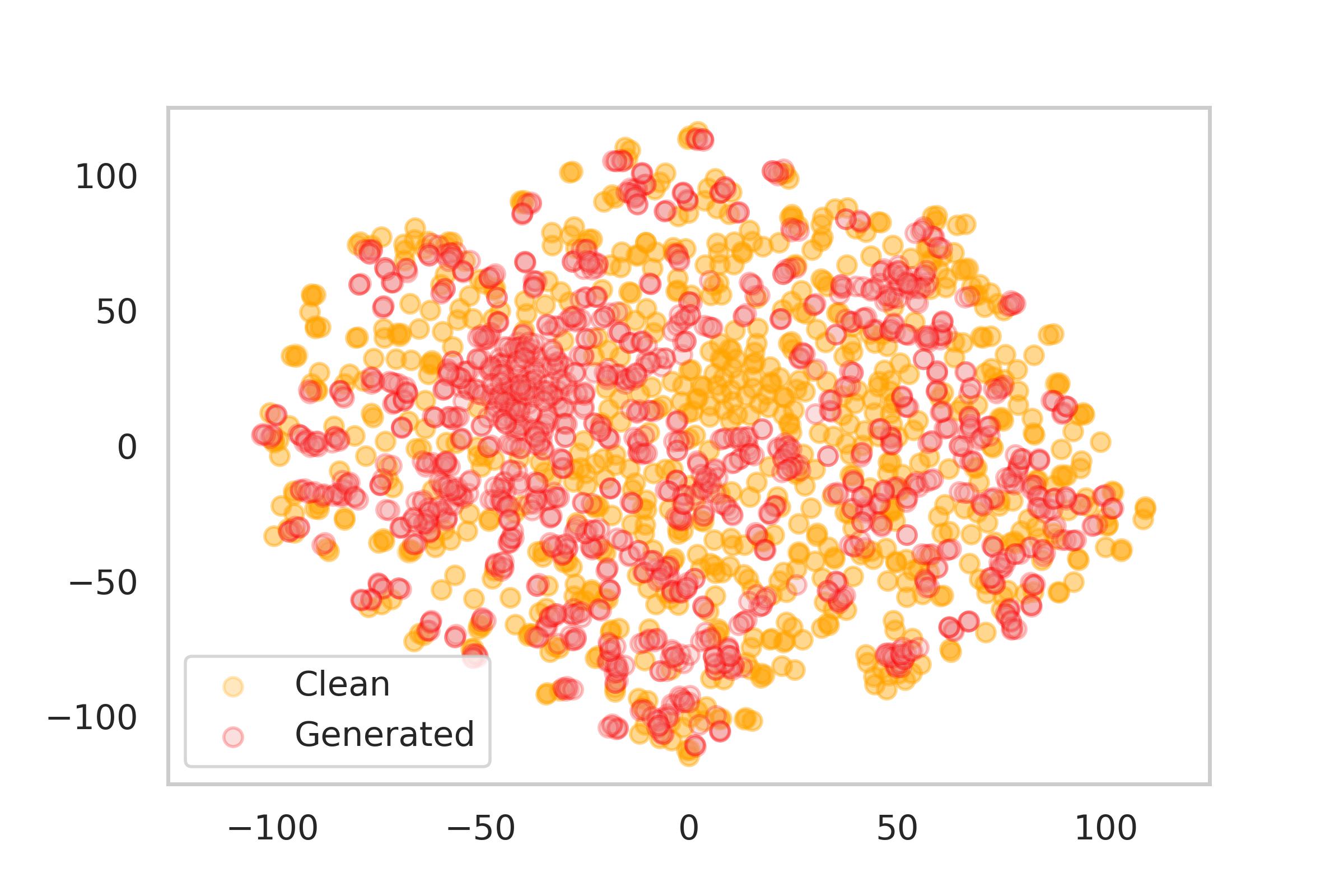}%
  \label{fig:tsne_visual}}
\caption{t-SNE visualization of the latent space for clean and polluted evidence across text and image modalities.}
\label{fig:tsne}
\end{figure}
% \endgroup

We randomly select  500 claims from the test set and visualize the 
similarity between their  polluted evidence and the original clean evidence.
 As shown in 
Figure \ref{fig:vis}, the distribution is centered around zero, indicating that the polluted evidence closely resembles the original clean evidence.
We apply t-SNE to visualize the latent spaces for clean and polluted textual and image evidence. Figure~\ref{fig:tsne} shows that our approach is able to generate evidence that not only closely mirrors the original clean evidence but also exhibits greater similarity to the input claim, thereby effectively contaminating the clean evidence while preserving high semantic similarity.

\subsection{Proposed Framework}

\begin{figure}[t!]
    \centering
    \includegraphics[width=\linewidth]{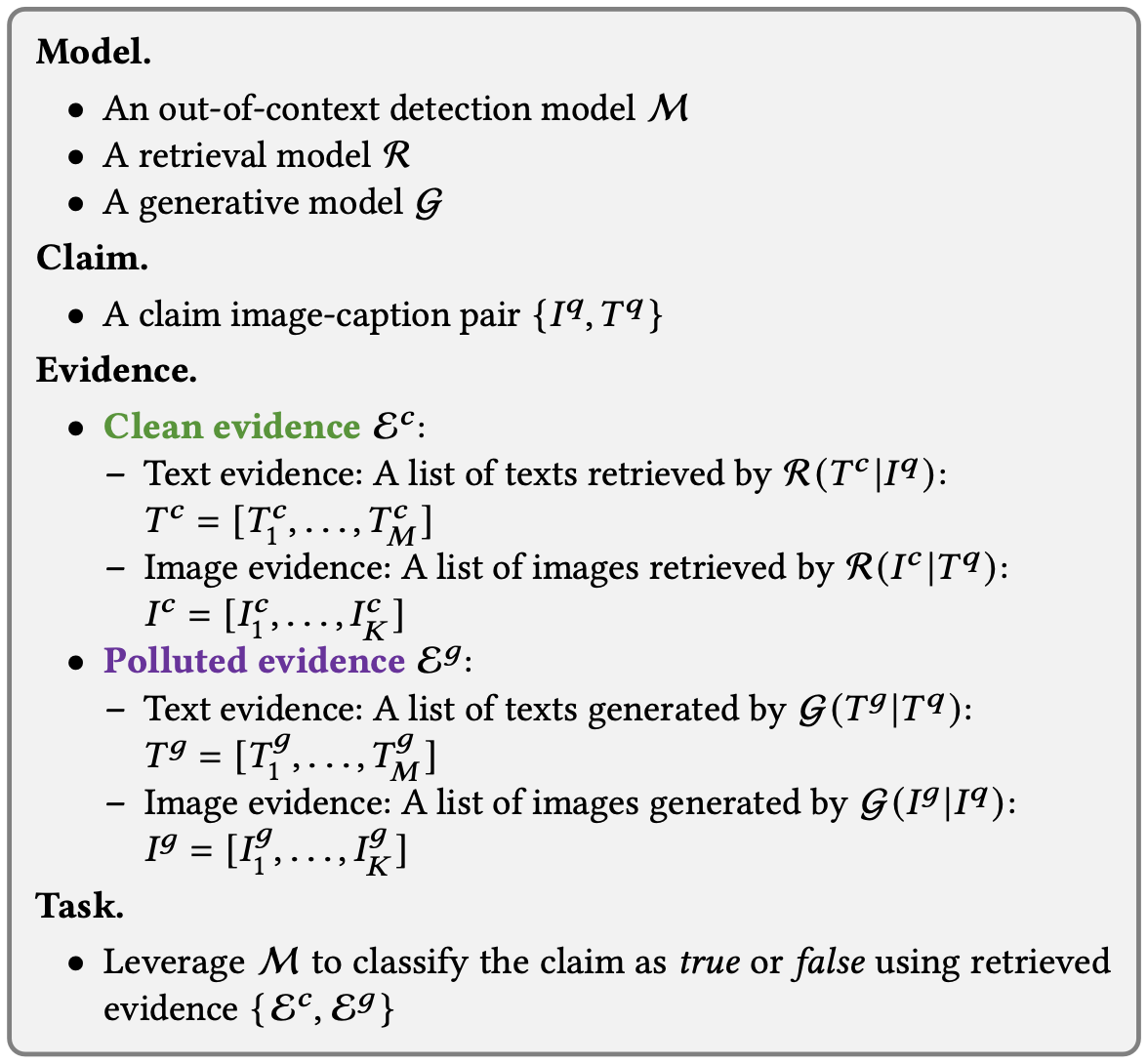}
    \caption{Notations and task definition.
    }
    \label{fig:task_formulation}
\end{figure}

\begin{figure*}[t!]
    \centering
    \includegraphics[width=\linewidth]{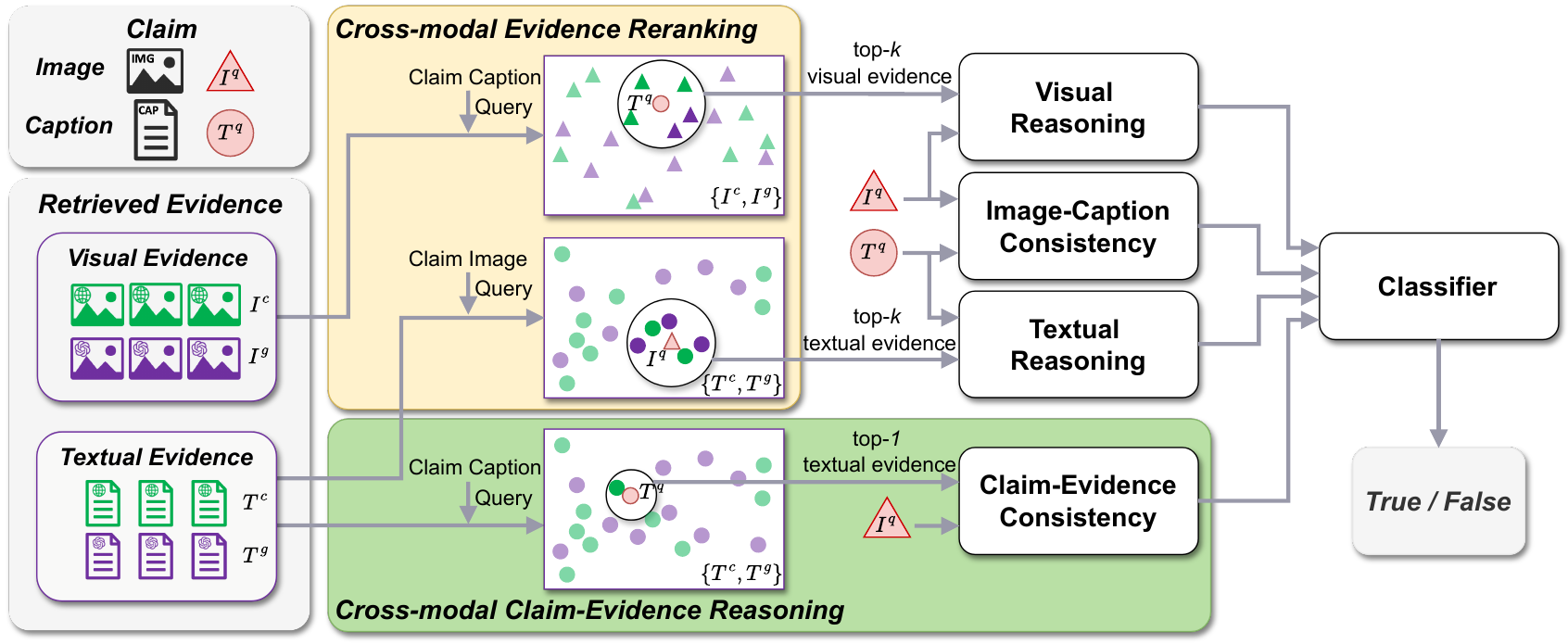}
    \caption{OOC misinformation detection framework in the presence of polluted evidence with proposed cross-modal reranking and claim-evidence reasoning. 
    }
    \label{fig:framework}
\end{figure*}

Figure~\ref{fig:task_formulation} summarize the notations used. An overview of our proposed framework is shown in Figure \ref{fig:framework}.
 Given a  claim comprising of an image $I^q$ and a caption $T^q$, 
 we first retrieve visual and textual evidence from the web using Google Vision
 and Google Custom Search.
 We introduce  a cross-modal re-ranking strategy to identify the top-k evidence before performing  
  visual, textual and image-caption consistency reasoning. 
  The visual reasoning module examines the relevance between the claim image  $I^q$ and polluted image evidence $\{I^c, I^g\}$. 
The textual reasoning module assesses how well the query caption $T^q$ corresponds with polluted text evidence $\{{T^c, T^g}\}$. Beyond these individual assessments, the consistency reasoning module checks the consistency between the claim image and the caption. 
 In addition, we also design a 
cross-modal claim-evidence reasoning strategy to ensure the consistency between claim caption and the top-1 textual evidence. 
The outputs from these reasoning modules are fused and passed to  a classifier to determine the claim's veracity.
Our proposed 
strategies enables us to
improve the robustness of OOC detectors in the presence of polluted evidence. We give the details 
below.

\subsection{Proposed Strategies}

\noindent\textbf{Cross-modal Evidence Re-ranking.} This strategy \textit{addresses the issue of OOC detectors that inadvertently focus on polluted evidence by giving priority to evidence that  best aligns with the claim}. 
Inspired by \cite{yao2023end}, 
 we use CLIP to identify the most contextually relevant textual evidence from a corpus that may contain  polluted information,  based on the claim image. Similarly, 
 this method is employed to determine the most relevant visual evidence based on the claim caption.
   Algorithm~\ref{alg:algorithm-1} gives details.

 Specifically, we 
 utilize CLIP embeddings to compute cross-modal similarity scores and
 obtain the re-ranked lists of visual and textual evidence. The top-k evidence are then passed to the visual and textual reasoning modules, respectively.

\smallskip
\noindent\textbf{Cross-modal Claim-Evidence Reasoning.}
Cross-modal claim-evidence reasoning  \textit{goes beyond traditional caption-image consistency check, which often misses critical contextual details} provided by external evidence. 
For example, a false caption may correctly describe the visible elements in an image but misrepresent its context, such as attributing a news event to the wrong location or time. These discrepancies can only be verified using external information that is most pertinent to the main entity in the caption.
As such, we use the most relevant textual evidence related to the caption for a consistency check with the claim image, ensuring the model's robustness even when confronted with polluted evidence. 
Algorithm~\ref{alg:algorithm-2} provides details.

\smallskip
\noindent
 The two proposed strategies can be utilized in a plug-and-play manner, allowing for easier integration into real-world applications, without the need for retraining. 
 Further, these strategies are adaptable to various types of pollution with the emphasis on enhancing semantic-level reasoning rather than overfitting to surface-level correlations or dataset-specific feature distributions introduced by a specific pollution model. \\

\begin{algorithm}[t!]
\caption{Cross-modal Evidence Reranking}
\label{alg:algorithm-1}
\textbf{Input}: claim $<$$I^q, T^q$$>$,  sets of retrieved textual evidence $\mathcal{T}=\{T^c, T^g\}$ and  
visual evidence $\mathcal{V}=\{I^c, I^g\}$\\
\textbf{Output}: sorted textual and visual evidences  \\ 
\begin{algorithmic}[1]
\STATE initialize $S_1 \gets [~]$, $S_2 \gets [~]$
\FOR{$T \in \mathcal{T}$}
    \STATE compute cross-modal similarity score
    \STATE $s \gets cos(\mbox{CLIP}(I^q), \mbox{CLIP}(T))$
    \STATE $S_1$.insert($s$)
\ENDFOR
\FOR{$V \in \mathcal{V}$}
    \STATE compute cross-modal similarity score
    \STATE $s \gets cos(\mbox{CLIP}(T^q), \mbox{CLIP}(V))$ 
    \STATE $S_2$.insert($s$)
\ENDFOR
\STATE return $\text{argsort}(S_1)$, $\text{argsort}(S_2)$ in descending order
\end{algorithmic}
\end{algorithm}

\begin{algorithm}[t!]
\caption{Cross-modal Claim-Evidence Reasoning}
\label{alg:algorithm-2}
\textbf{Input}: claim $<$$I^q, T^q$$>$, set of retrieved textual evidence $\mathcal{T}$, claim-evidence consistency reasoning module $\mathcal{M}$\\
\textbf{Output}: reasoning-representation\\
\begin{algorithmic}[1]
\STATE Initialize $S \gets [~]$.
\FOR{$T \in \mathcal{T}$}
    \STATE compute intra-modal similarity score
    \STATE $s \gets cos(\mbox{CLIP}(T^q),\mbox{CLIP}(T))$ 
    \STATE $S$.insert($s$) 
\ENDFOR
\STATE Return  $\mathcal{M} (\mathcal{T}[\text{argmax}(S)$], $I^q)$
\end{algorithmic}
\end{algorithm}

\section{Performance Study}

 In this section, we show the performance degradation of existing OOC detectors in the presence of polluted evidence. Then we demonstrate the effectiveness of our proposed strategies in mitigating the effect of polluted evidence.

\smallskip
\noindent\textbf{Datasets.}
We use the following datasets in our experiments:
\begin{itemize}
\item \textbf{NewsCLIPpings} \cite{luo-etal-2021-newsclippings} is the largest synthetic benchmark for OOC misinformation detection. 
It synthesizes  out-of-context samples by replacing the images in the original image-caption pairs with retrieved images that are semantically related but belong to different news events. \cite{abdelnabi2022open} extends this dataset by supplementing both textual and visual evidence using Google Search APIs.

\item \textbf{VERITE} \cite{papadopoulos2024verite} is a real-world benchmark for evaluating multimodal misinformation detection. It consists of real and out-of-context pairs collected from fact-checking websites, such as Snopes
and Reuters.
We use the corresponding multimodal evidence from \cite{papadopoulos2023reddot}. 

\end{itemize}

\smallskip
\noindent For each piece of textual evidence, we randomly apply one of the LLM instructions to create the corresponding polluted entity-based, supporting or  refuting evidence via GPT-4 \cite{openai2024gpt4}. 
For each piece of visual evidence, we use Depth-conditioned Stable Diffusion \cite{Rombach_2022_CVPR} to generate the corresponding images.  
The polluted evidence is added to the original clean evidence corpus. 
Table \ref{tab:dataset} shows the statistics 
for the two datasets.

\begin{table}[h!]
\caption{Dataset Statistics}
\begin{center}
\begin{tabular}{@{}lcccc@{}}
\toprule
\multirow{2}{*}{\textbf{Dataset}} & \multicolumn{3}{c}{\textbf{NewsCLIPpings}} & \multicolumn{1}{c}{\textbf{VERITE}} \\
\cmidrule(lr){2-4} \cmidrule(lr){5-5}
& Train & Validation & Test & Test \\
\midrule
\textbf{\textit{Claim}} & 71,072 & 7,024 & 7,264 & 662 \\
\midrule
\multirow{1}{*}{\textbf{\textit{Evidence}}}  & & & &\\
Clean Text & 689,995 & 58,388 & 60,848 & 1,261 \\
Polluted Text & 903,067 & 82,112 & 67,016 & 2,002 \\
Clean Image  & 650,738 & 64,562 & 66,772 & 8,309 \\
Polluted Image  & 655,848 & 65,082 & 67,092 & 8,389 \\
\bottomrule
\end{tabular}
\label{tab:dataset}
\end{center}
\end{table}

\noindent\textbf{Baselines.} We use the following OOC detectors:

\begin{itemize}

 \item \textbf{CCN} \cite{abdelnabi2022open}. This employs attention-based memory networks for visual and textual reasoning between the claim and evidence, and a fine-tuned CLIP 
component to check the claim image and caption consistency.

 \item \textbf{RED-DOT} \cite{papadopoulos2023reddot}.
This leverages the pre-trained CLIP
as the backbone to extract visual and textual features. A transformer-based  fusion module is used to facilitate interaction and reasoning among these features.

 \item \textbf{SNIFFER} \cite{qi2024sniffer}.  This is the state-of-the-art multimodal large language model designed for OOC misinformation detection. It employs a two-stage instruction tuning on InstructBLIP
for the cross-modal consistency checks.

 \item \textbf{GPT-4o} \cite{openai2024gpt4o}. This is currently one of the most powerful multimodal large language models. 
We utilize GPT-4o in a zero-shot manner with step-by-step instructions for OOC misinformation detection. \\

\end{itemize}

\noindent\textbf{Implementation Details}
We use CCN \cite{abdelnabi2022open} and SNIFFER's \cite{qi2024sniffer} public model checkpoints fine-tuned on the NewsCLIPpings training set. 
We use the InstructBLIP \cite{dai2023instructblip} as our captioner for visual reasoning path 
without fine-tuning. We leverage the CLIP (\verb|ViT-L/14|) as the cross-modal reranking module and select the top-1 sentence and top-5 images for textual and visual evidence. For augmented reasoning, we reuse the original CLIP component from CCN and internal checking from SNIFFER. All models are trained and evaluated on 8 Nvidia H100 (80G) GPUs. Our two proposed strategies are applied during inference and do not require any additional training. We generate textual pollution with GPT-4 (\verb|gpt-4|) \cite{openai2024gpt4}, which is configured with a temperature of 1.2, a maximum token length of 64, and a top-P setting of 0.95. We employ the variant of Stable Diffusion v2 models (
\texttt{stabilityai/\allowbreak stable-diffusion-\allowbreak 2-depth}
) to generate visual pollution. 
We report accuracy over all samples and F1 scores for the true and false samples, respectively.

\subsection{Effect of Evidence Pollution on OOC Detectors}
\label{sec:experiment-pollution}

Table \ref{tab:pollution} shows the 
OOC detection performance across different evidence  modalities.
We observe that: 
\begin{itemize}
    \item The combination of polluted text and image  poses a significant threat to OOC detectors. Specifically, the accuracy of all OOC  detectors drops by more than 9 percentage points, revealing the vulnerabilities of existing OOC detectors against generated multimodal pollution. 

    \item Textual pollution has a greater impact than  visual pollution, indicating that existing OOC detectors are more dependent on textual information. This modality bias may stem from the fact that textual evidence often provides more semantics such as relationships between entities compared to images. 

    \item Detection of false claims  in the presence of polluted evidence proves to be more challenging than detecting true claims. Specifically, CCN experiences a significant drop of 35.67 points in the F1 score for false claims on the VERITE dataset, highlighting the difficulties in reasoning with contradictory evidence.
\end{itemize}

In addition to the closed-source GPT-4o, we also evaluate open-source LLM-based detectors. As shown in Table~\ref{tab:more_models}, both LLaVA and InstructBLIP exhibit substantial performance drops when exposed to multimodal polluted evidence, revealing their vulnerability to AI-generated multimodal pollution.

\begin{table*}[t!]
\centering
% \small 
\caption{OOC detection performance (\%) under evidence pollution of different modalities. The first row (Clean) refers to the original performance without any pollution. The absolute change compared to the Clean setting is highlighted in \color{red}{red}.}
\label{tab:pollution}
\begin{NiceTabular}{llllllll}
\CodeBefore
    \rectanglecolor{gray!10}{4-2}{2-8}
    \rectanglecolor{gray!10}{8-2}{6-8}
    \rectanglecolor{gray!10}{12-2}{10-8}
    \rectanglecolor{gray!10}{16-2}{14-8}
\Body 
\toprule

\multirow{2}{*}{} & \multirow{2}{*}{\textbf{{Evidence}}} & \multicolumn{3}{c}{\textbf{NewsCLIPpings}} & \multicolumn{3}{c}{\textbf{VERITE}} \\ \cmidrule(lr){3-5} \cmidrule(lr){6-8}  
    &                   & \multicolumn{1}{c}{\textbf{Acc.}}      & \multicolumn{1}{c}{\textbf{F1-True}} & \multicolumn{1}{c}{\textbf{F1-False}}     & \multicolumn{1}{c}{\textbf{Acc.}}        & \multicolumn{1}{c}{\textbf{F1-True}} & \multicolumn{1}{c}{\textbf{F1-False}}  \\ 
\midrule

\parbox[t]{2mm}{\multirow{4}{*}{\rotatebox[origin=c]{90}{CCN}}}  &  Clean        &  84.28   &  84.29 & 84.27  &  67.25  &  71.52 &   61.48 \\
& Polluted Text  & 75.12 \textcolor{red}{\scriptsize ($\downarrow$9.16)}   & 78.10 \textcolor{red}{\scriptsize ($\downarrow$6.19)}  & 71.22 \textcolor{red}{\scriptsize ($\downarrow$13.05)} & 59.06 \textcolor{red}{\scriptsize ($\downarrow$8.19)} & 69.91 \textcolor{red}{\scriptsize ($\downarrow$1.61)}   & 35.97 \textcolor{red}{\scriptsize ($\downarrow$25.51)}  \\
& Polluted Image  & 82.11 \textcolor{red}{\scriptsize ($\downarrow$2.17)}   & 82.85 \textcolor{red}{\scriptsize ($\downarrow$1.44)}  & 81.30 \textcolor{red}{\scriptsize ($\downarrow$2.97)} & 63.41 \textcolor{red}{\scriptsize ($\downarrow$3.84)} & 68.93 \textcolor{red}{\scriptsize ($\downarrow$2.59)}  & 55.51 \textcolor{red}{\scriptsize ($\downarrow$5.97)}  \\
& Polluted Text + Image  & 71.78 \textcolor{red}{\scriptsize ($\downarrow$12.50)}   & 76.48 \textcolor{red}{\scriptsize ($\downarrow$7.81)}  & 64.72 \textcolor{red}{\scriptsize ($\downarrow$19.55)} & 55.92 \textcolor{red}{\scriptsize ($\downarrow$11.33)} & 68.65 \textcolor{red}{\scriptsize ($\downarrow$2.87)}  & 25.81 \textcolor{red}{\scriptsize ($\downarrow$35.67)}  \\
\midrule
\parbox[t]{2mm}{\multirow{4}{*}{\rotatebox[origin=c]{90}{RED-DOT}}}  & Clean        & 84.98   & 84.62 & 85.32  & 64.29  & 62.39 & 66.00 \\ 
& Polluted Text  & 75.56 \textcolor{red}{\scriptsize ($\downarrow$9.42)}   & 70.62 \textcolor{red}{\scriptsize ($\downarrow$14.00)}  & 79.09 \textcolor{red}{\scriptsize ($\downarrow$6.23)} & 52.64 \textcolor{red}{\scriptsize ($\downarrow$11.65)} & 50.73 \textcolor{red}{\scriptsize ($\downarrow$11.66)} & 54.24 \textcolor{red}{\scriptsize ($\downarrow$11.76)} \\
& Polluted Image  & 79.85 \textcolor{red}{\scriptsize ($\downarrow$5.13)}   & 76.81 \textcolor{red}{\scriptsize ($\downarrow$7.81)}  & 82.19 \textcolor{red}{\scriptsize ($\downarrow$3.13)} & 57.49 \textcolor{red}{\scriptsize ($\downarrow$6.80)} & 57.93 \textcolor{red}{\scriptsize ($\downarrow$4.46)} & 57.04 \textcolor{red}{\scriptsize ($\downarrow$8.96)} \\
& Polluted Text + Image  & 73.75 \textcolor{red}{\scriptsize ($\downarrow$11.23)}   & 67.19 \textcolor{red}{\scriptsize ($\downarrow$17.43)} & 78.12 \textcolor{red}{\scriptsize ($\downarrow$7.20)} & 48.75 \textcolor{red}{\scriptsize ($\downarrow$15.54)} & 48.65 \textcolor{red}{\scriptsize ($\downarrow$13.74)} & 48.85 \textcolor{red}{\scriptsize ($\downarrow$17.15)} \\
\midrule
\parbox[t]{2mm}{\multirow{4}{*}{\rotatebox[origin=c]{90}{SNIFFER}}}  & Clean        & 88.85   & 88.92 & 88.78  & 73.69  & 76.15 & 70.68 \\ 
& Polluted Text  & 78.55 \textcolor{red}{\scriptsize ($\downarrow$10.30)}   & 80.08 \textcolor{red}{\scriptsize ($\downarrow$8.84)}  & 77.21 \textcolor{red}{\scriptsize ($\downarrow$11.57)} & 65.16 \textcolor{red}{\scriptsize ($\downarrow$8.53)} & 68.75 \textcolor{red}{\scriptsize ($\downarrow$7.40)} & 60.99 \textcolor{red}{\scriptsize ($\downarrow$9.69)} \\
& Polluted Image  & 82.25 \textcolor{red}{\scriptsize ($\downarrow$6.60)}   & 82.19 \textcolor{red}{\scriptsize ($\downarrow$6.73)}  & 82.50 \textcolor{red}{\scriptsize ($\downarrow$6.28)} & 67.94 \textcolor{red}{\scriptsize ($\downarrow$5.75)} & 71.13 \textcolor{red}{\scriptsize ($\downarrow$5.02)} & 64.48 \textcolor{red}{\scriptsize ($\downarrow$6.20)} \\
& Polluted Text + Image  & 76.42 \textcolor{red}{\scriptsize ($\downarrow$12.43)}   & 77.47 \textcolor{red}{\scriptsize ($\downarrow$11.45)} & 75.31 \textcolor{red}{\scriptsize ($\downarrow$13.47)} & 59.41 \textcolor{red}{\scriptsize ($\downarrow$14.28)} & 64.71 \textcolor{red}{\scriptsize ($\downarrow$11.44)} & 53.04 \textcolor{red}{\scriptsize ($\downarrow$17.64)} \\
\midrule
\parbox[t]{2mm}{\multirow{4}{*}{\rotatebox[origin=c]{90}{GPT-4o}}}  & Clean        & 87.27   & 86.58 & 87.89  & 77.53  & 76.76 & 78.25 \\ 
& Polluted Text  & 79.02 \textcolor{red}{\scriptsize ($\downarrow$8.25)}   & 75.88 \textcolor{red}{\scriptsize ($\downarrow$10.70)}  & 81.44 \textcolor{red}{\scriptsize ($\downarrow$6.45)} & 67.42 \textcolor{red}{\scriptsize ($\downarrow$10.11)} & 63.12 \textcolor{red}{\scriptsize ($\downarrow$13.64)} & 70.83 \textcolor{red}{\scriptsize ($\downarrow$7.42)} \\
& Polluted Image  & 82.48 \textcolor{red}{\scriptsize ($\downarrow$4.79)}   & 81.44 \textcolor{red}{\scriptsize ($\downarrow$5.14)}  & 83.41 \textcolor{red}{\scriptsize ($\downarrow$4.48)} & 68.64 \textcolor{red}{\scriptsize ($\downarrow$8.89)} & 65.91 \textcolor{red}{\scriptsize ($\downarrow$10.85)} & 70.97 \textcolor{red}{\scriptsize ($\downarrow$7.28)} \\
& Polluted Text + Image  & 77.72 \textcolor{red}{\scriptsize ($\downarrow$9.55)}   & 74.69 \textcolor{red}{\scriptsize ($\downarrow$11.89)} & 80.11 \textcolor{red}{\scriptsize ($\downarrow$7.78)} & 64.29 \textcolor{red}{\scriptsize ($\downarrow$13.24)} & 56.29 \textcolor{red}{\scriptsize ($\downarrow$20.47)} & 69.81 \textcolor{red}{\scriptsize ($\downarrow$8.44)} \\
\bottomrule
\end{NiceTabular}
\end{table*}

\begin{table}[t!]
  \caption{OOC detection performance (\%) of open-source LLM-based detectors under evidence pollution on NewsCLIPpings.}
  \centering
  \begin{tabular}{lcc}
    \toprule
    Model & Evidence & Acc. \\
    \midrule
    \multirow{2}{*}{LLaVA \cite{li2024llava}} & Clean & 57.00 \\
                           & Polluted Text + Image & 49.50 \\
    \midrule
    \multirow{2}{*}{InstructBLIP \cite{dai2023instructblip}} & Clean & 63.16 \\
                                  & Polluted Text + Image & 45.09 \\
    \bottomrule
  \end{tabular}
\label{tab:more_models}
\end{table}

\begin{figure}[t!]
\centering

  \includegraphics[width=0.9\linewidth]{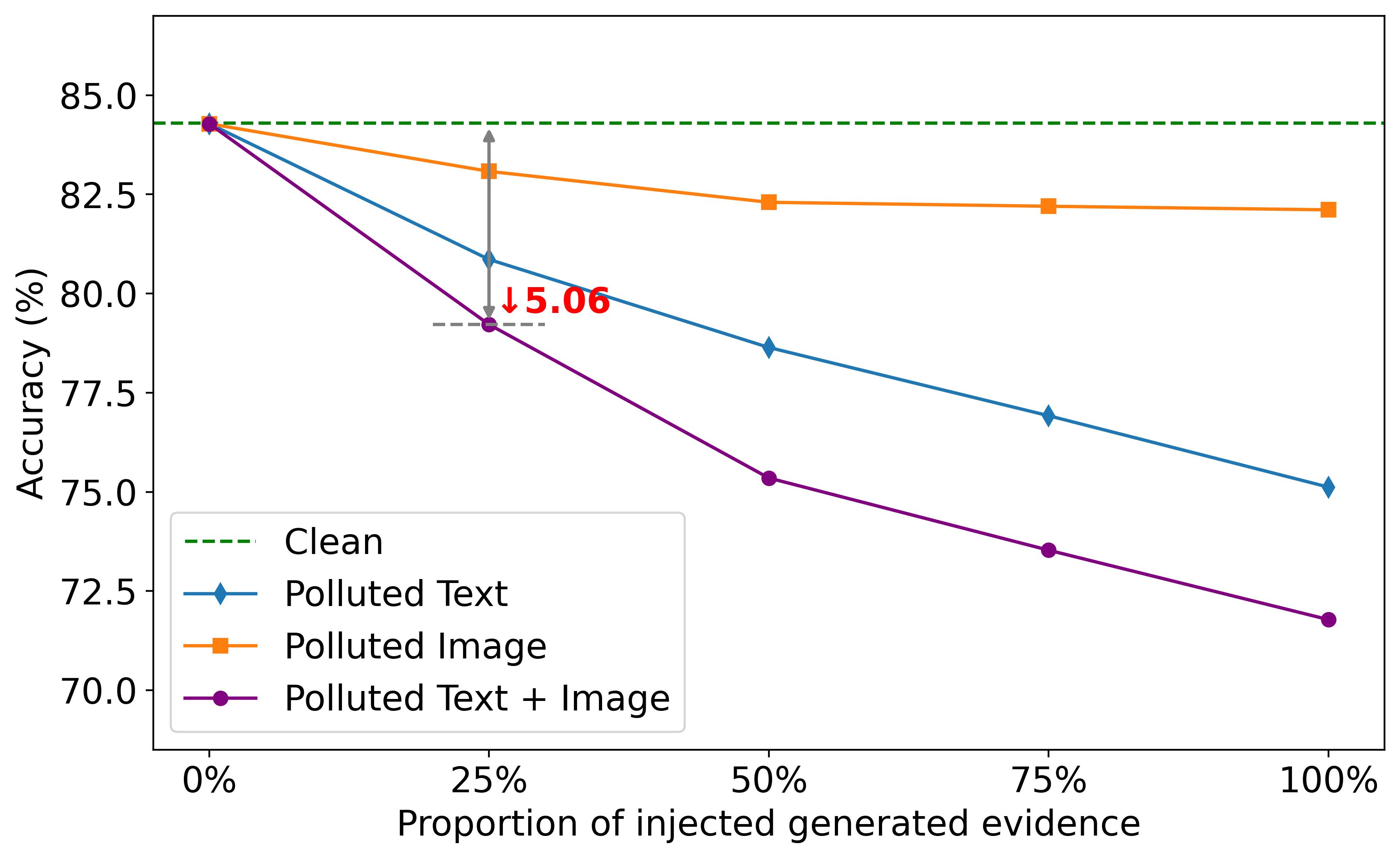}
 
  (a) CCN\\

  \includegraphics[width=0.9\linewidth]{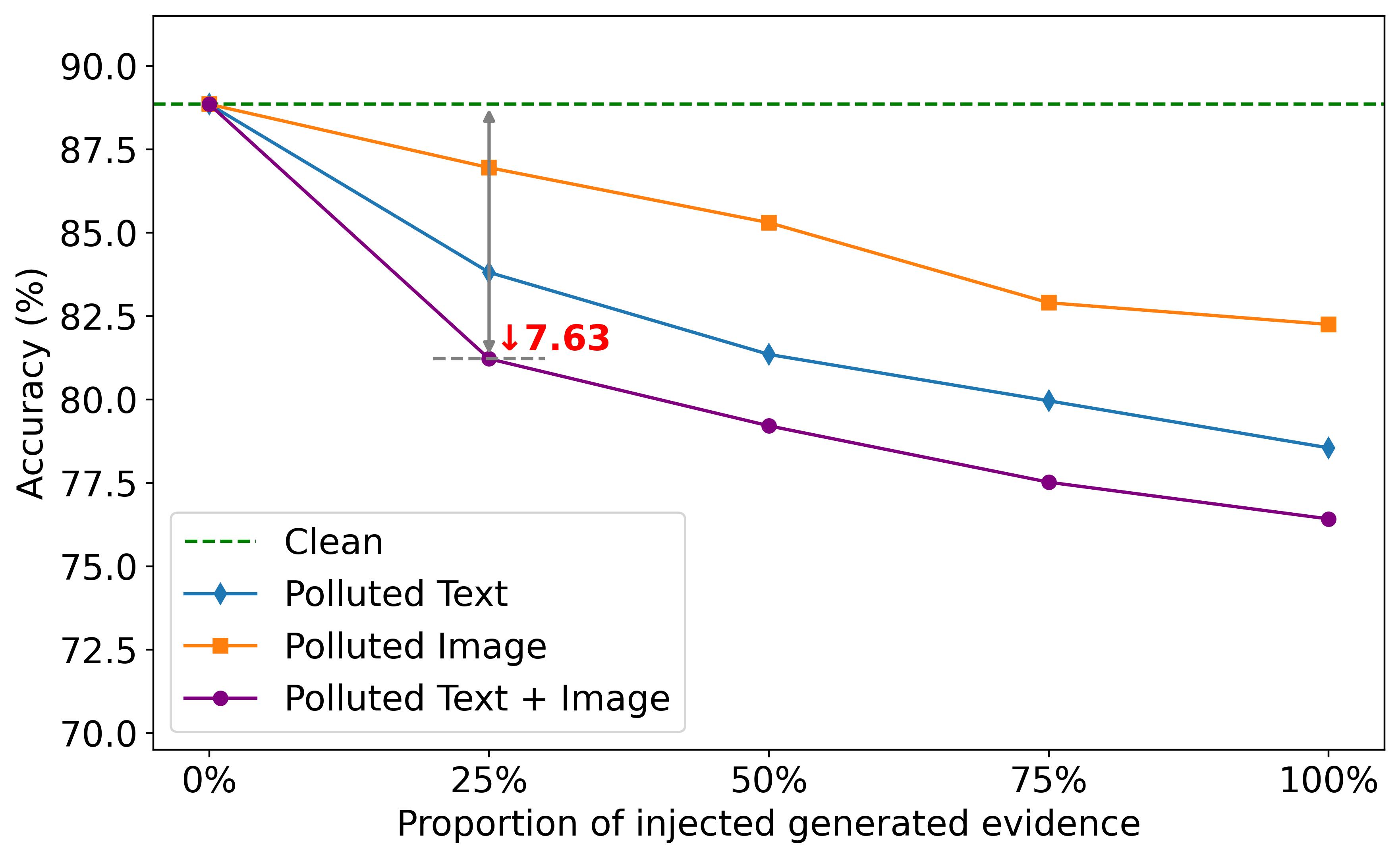}
  
(b) SNIFFER

\caption{Performance across varying proportion of polluted evidence on NewsCLIPpings dataset.}
\label{fig:inject_ratio}
\end{figure}

\smallskip
\noindent\textbf{Quantitative Analysis.} Figure \ref{fig:inject_ratio} 
shows the performance of CCN and SNIFFER when we vary the proportion of polluted evidence. We see that the accuracy of both models drops as the proportion of pollution increases. Even a small amount of pollution can significantly affect the model's detection capabilities. For instance, introducing  25\%  of polluted evidence results in a decrease of 5.06 points in the accuracy of CCN  and 7.63 points in SNIFFER. 
The accuracy of CCN demonstrates a noticeable decline as the level of evidence pollution increases, highlighting CCN's heavy reliance on the text modality for misinformation identification, making it particularly vulnerable to pollution introduced by LLMs.

\begin{figure}[t!]
\centering
\includegraphics[width=0.95\linewidth]{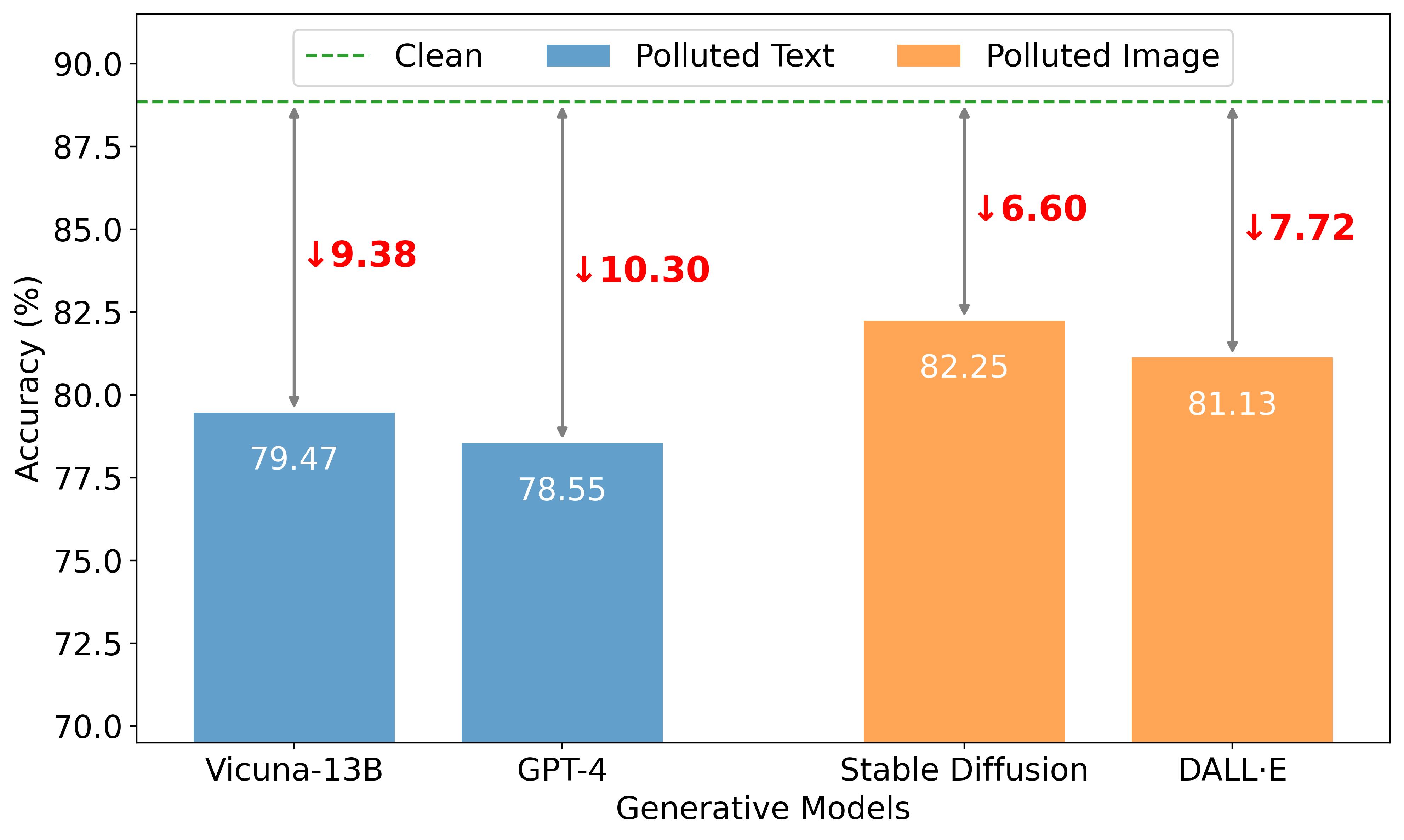}
\caption{SNIFFER's performance across different GenAI-based pollution models on NewsCLIPpings dataset.}
\label{fig:exp_genai}
\end{figure}

\smallskip
\noindent\textbf{Generalization Analysis.} Figure \ref{fig:exp_genai} shows the impact of pollution in textual and visual modalities under different generative models. Notably,  advanced models like DALL-E, which significantly improves image quality and resolution, further amplify the effects of visual pollution.
We also evaluate the performance of SNIFFER on two real world OOC datasets: TamperedNews-Ent \cite{sahar2025verifying}, utilizing English news from the BreakingNews \cite{RamisaYMM18breakingnews}, and News400-Ent \cite{sahar2025verifying}, which covers diverse domains (politics, economics, sports and travel) in German. Table~\ref{tab:more_real_data} shows that the accuracy of SNIFFER dropped by 10.55\% on TamperedNews-Ent and by 7.36\% on News400-Ent, highlighting 
the impact of GenAI-polluted evidence across different domains and languages.

\smallskip
\noindent\textbf{Human Evaluation.} 
To better assess the effectiveness of evidence pollution, we conducted a human evaluation on ten randomly selected misinformation samples with polluted evidence. Each sample includes multiple sub-questions for both clean and polluted evidence, resulting in a total of 100 evaluation questions. Twenty participants were asked to judge each piece of evidence's authenticity and each claim's veracity before and after reading the polluted evidence. 
The results show that (a) only 49.39\% of the polluted evidence was correctly identified as AI-generated, suggesting that nearly half of the polluted samples were perceived as authentic; 
(b) among these, textual pollution appeared more deceptive, with only 29.83\% of LLM-generated text pieces correctly identified by participants, compared to 73.79\% for visual pollution;
(c) 41.84\% of initially correct veracity judgments for misinformation samples were reversed to wrong predictions after reading polluted evidence,  highlighting the substantial impact of AI-generated content on human decision-making.

\subsection{Effect of Proposed Strategies}
\label{sec:experiment-strategy}

 We compare our strategies  with  baselines adapted from \cite{pan-etal-2023-risk}: 
\begin{itemize}
    \item \noindent\textbf{Source Filtering.} We check the source against a list of reputable websites. Specifically, we filter the retrieved evidence by retaining only those documents originating from domains listed in the top-1 million most popular websites based on Tranco \cite{LePochat2019}.

    \item \noindent\textbf{Extra Detector.} We introduce an auxiliary verifier that checks retrieved evidence before reasoning. Specifically, we fine-tune a lightweight CLIP-based classifier to distinguish between real and polluted evidence. We use 10,000 randomly sampled instances from the polluted corpus of NewsCLIPpings for training and 1,000 instances for validation, which contains genuine news claims as well as synthetic items generated with GPT-4 and Stable Diffusion model. This \q{filter-before-reason} design reduces direct exposure to polluted inputs at the cost of additional detector training.

    \item \noindent\textbf{Vigilant Prompting.} LLMs have demonstrated strong instruction-following capabilities when guided by appropriately designed prompts \cite{DBLP:conf/nips/Ouyang0JAWMZASR22}.  Without changing the architecture or re-training, we inject explicit cautionary instructions into the prompt (e.g., \q{Some sentences in the evidence may be polluted to mislead you. You should verify credibility before answering.}). 

    \item \noindent\textbf{Reader Ensemble.} This method applies a divide-and-vote scheme: retrieved evidence passages are randomly partitioned into $k$ groups, the reader produces $k$ answers, and the final decision $a_v$ is obtained by majority voting over the $k$ candidate responses $a_1, ..., a_k \in\{\text{True},\text{False}\}$. This voting strategy aims to minimize the impact of misinformation by limiting the influence of any single contaminated information sources on final prediction.
    
\end{itemize}

As shown in Table \ref{tab:strategy_comparison}, SNIFFER equipped with our proposed strategies achieves the highest performance across two datasets, with significant improvements of 12.40\% on NewsCLIPpings and 13.41\% on VERITE,
demonstrating its superiority in the presence of polluted evidence. 
In contrast, source filtering fails to filter out pollution from high-traffic user-generated content platforms (e.g., social media, forums) which are also ranked highly on Tranco. Furthermore, strict allowlisting inadvertently discards valid long-tail evidence from smaller local news outlets, reducing the critical semantic context for verification.

Note that our approaches can be easily incorporated into existing OOC detection frameworks, such as CCN and RED-DOT, whereas the prompting-based and voting-based approaches are restricted to LLM-based detectors. 
Combining our strategies with vigilant prompting further boosts SNIFFER's performance, achieving cumulative improvements of 13.43\% on NewsCLIPpings and 14.86\% on VERITE, highlighting the flexibility of our proposed strategies.

\begin{table}[t!]
  \caption{SNIFFER's performance (\%) on two real-world datasets.}
  \centering
  \begin{tabular}{lcc}
    \toprule
    Dataset & Evidence & Acc. \\
    \midrule
    \multirow{2}{*}{TamperedNews-Ent \cite{sahar2025verifying}} & Clean & 85.19 \\
                           & Polluted Text + Image & 74.64 \\
    \midrule
    \multirow{2}{*}{News400-Ent \cite{sahar2025verifying}} & Clean & 80.06 \\
                                  & Polluted Text + Image & 72.70 \\
    \bottomrule
  \end{tabular}
\label{tab:more_real_data}
\end{table}

\begin{table*}[t!]
\centering
\caption{Performance comparison of different strategies. The first row (None) refers to the original performance under multimodal evidence pollution.}
\label{tab:strategy_comparison}
\begin{NiceTabular}{llllllll}
\CodeBefore
    \rectanglecolor{gray!10}{4-2}{2-8}
    \rectanglecolor{gray!10}{9-2}{7-8}
\Body 
\toprule
 \multirow{2}{*}{} & \multirow{2}{*}{\textbf{{Strategy}}} & \multicolumn{3}{c}{\textbf{NewsCLIPpings}} & \multicolumn{3}{c}{\textbf{VERITE}} \\ \cmidrule(lr){3-5} \cmidrule(lr){6-8}  
    &                   & \multicolumn{1}{c}{\textbf{Acc.}}      & \multicolumn{1}{c}{\textbf{F1-True}} & \multicolumn{1}{c}{\textbf{F1-False}}     & \multicolumn{1}{c}{\textbf{Acc.}}        & \multicolumn{1}{c}{\textbf{F1-True}} & \multicolumn{1}{c}{\textbf{F1-False}}  \\ 
\midrule
\multirow{5}{*}{\rotatebox[origin=c]{90}{SNIFFER}}  
& None                              & 76.42               & 77.47                   & 75.31     &59.41&64.71&53.04              \\ 
&Source Filtering                & 73.71 (↓2.71)       & 77.82 (↑0.35)           & 70.10 (↓5.21)&60.18 (↑0.77)&65.74 (↑1.03)& 54.92 (↑1.88)          \\ 
&Extra Detector                & 79.00 (↑2.58)       & 80.73 (↑3.26)           & 76.92 (↑1.61)&68.99 (↑9.58)&72.01 (↑7.30)& 65.23 (↑12.19)          \\ 
&Vigilant Prompting               & 79.49 (↑3.07)       & 80.93 (↑3.46)           & 77.84 (↑2.53)   &69.51 (↑10.10)&72.18 (↑7.47)&66.28 (↑13.24)        \\ 
&Reader Ensemble                 & 68.51 (↓7.91)       & 70.70 (↓6.77)           & 65.94 (↓9.37)    &64.81 (↑5.40)& 63.54 (↓1.17)& 65.99 (↑12.95)      \\ 
\RowStyle{\bfseries}&Ours                         & 88.82 (↑12.40)      & 89.15 (↑11.68)          & 88.48 (↑13.17)   &72.82 (↑13.41)&76.00 (↑11.29)&68.67 (↑15.63)       \\ 
 \midrule
 \multirow{5}{*}{\rotatebox[origin=c]{90}{GPT-4o}}  
 & None                              & 77.72               & 74.69                   & 80.11     &64.29&56.29&69.81              \\ 
 &Source Filtering                & 76.90 (↓0.82)       & 74.32 (↓0.37)           & 79.76 (↓0.35)&64.56 (↑0.27)&57.13 (↑0.84)& 72.51 (↑2.70)          \\ 
 &Extra Detector                &  81.69 (↑3.97)       & 79.20 (↑4.51)           & 83.85 (↑3.74)&72.13 (↑7.84)&70.15 (↑13.86) & 74.10 (↑4.29)          \\ 
 &Vigilant Prompting               & 83.50 (↑5.78)       & 82.50 (↑7.81)           & 84.84 (↑4.73)   &66.03 (↑1.74)&62.14 (↑5.85) & 69.41 (↓0.40)       \\ 
 &Reader Ensemble                 & 72.33 (↓5.39)       & 68.53 (↓6.16)           & 77.29 (↓2.82)    &64.98 (↑0.69)&62.20 (↑5.91)&68.80 (↓1.01)       \\ 
 \RowStyle{\bfseries}&Ours                         & 88.00 (↑10.28)      & 87.51 (↑12.82)          & 88.53 (↑8.42)   &75.44 (↑11.15)&76.30 (↑20.01)&74.50 (↑4.69)       \\ 
\bottomrule
\end{NiceTabular}
\end{table*}

\begin{table*}[t!]
\centering
\caption{
OOC detection performance (\%) with the proposed strategies under the evidence pollution. The first row (None) refers to the original performance under multimodal pollution. The absolute change to the original one is highlighted in \color{blue}{blue}.}
\label{tab:defense}
\begin{NiceTabular}{llllllll}
\CodeBefore
    \rectanglecolor{gray!10}{4-2}{2-8}
    \rectanglecolor{gray!10}{8-2}{6-8}
    \rectanglecolor{gray!10}{12-2}{10-8}
    \rectanglecolor{gray!10}{16-2}{14-8}
\Body 
\toprule
 \multirow{2}{*}{} & \multirow{2}{*}{\textbf{{Strategy}}} & \multicolumn{3}{c}{\textbf{NewsCLIPpings}} & \multicolumn{3}{c}{\textbf{VERITE}} \\ \cmidrule(lr){3-5} \cmidrule(lr){6-8}  
    &                   & \multicolumn{1}{c}{\textbf{Acc.}}      & \multicolumn{1}{c}{\textbf{F1-True}} & \multicolumn{1}{c}{\textbf{F1-False}}     & \multicolumn{1}{c}{\textbf{Acc.}}        & \multicolumn{1}{c}{\textbf{F1-True}} & \multicolumn{1}{c}{\textbf{F1-False}}  \\ 
\midrule
\multirow{4}{*}{\rotatebox[origin=c]{90}{CCN}}  
&    None    & 71.78 & 76.48 &64.72 & 55.92 & 68.65 &25.81\\ 
& Cross-modal Reranking   & 79.70 \textcolor{blue}{\scriptsize ($\uparrow$7.92)}  & 79.88 \textcolor{blue}{\scriptsize ($\uparrow$3.40)}  & 79.51 \textcolor{blue}{\scriptsize ($\uparrow$14.79)} & 61.67 \textcolor{blue}{\scriptsize ($\uparrow$5.75)}  & 65.08 \color{black}{\scriptsize ($\downarrow$3.57)}  & 57.53 \textcolor{blue}{\scriptsize ($\uparrow$31.72)} \\
& Cross-modal Reasoning   & 75.17 \textcolor{blue}{\scriptsize ($\uparrow$3.39)}  & 78.38 \textcolor{blue}{\scriptsize ($\uparrow$1.90)}  & 70.83 \textcolor{blue}{\scriptsize ($\uparrow$6.11)}  & 59.76 \textcolor{blue}{\scriptsize ($\uparrow$3.84)}  & 63.39 \color{black}{\scriptsize ($\downarrow$5.26)}  & 55.32 \textcolor{blue}{\scriptsize ($\uparrow$29.51)}\\
&  \makecell[cl]{Both}   & 80.21 \textcolor{blue}{\scriptsize ($\uparrow$8.43)}  & 80.86 \textcolor{blue}{\scriptsize ($\uparrow$4.38)}  & 79.52 \textcolor{blue}{\scriptsize ($\uparrow$14.80)} & 65.51 \textcolor{blue}{\scriptsize ($\uparrow$9.59)}  & 70.54 \textcolor{blue}{\scriptsize ($\uparrow$1.89)}  & 58.40 \textcolor{blue}{\scriptsize ($\uparrow$32.59)} \\
\midrule
\multirow{4}{*}{\rotatebox[origin=c]{90}{RED-DOT}}  &   None     & 73.75 & 67.19 &78.12& 48.75 & 48.65 & 48.85\\ 
& Cross-modal Reranking   & 82.92 \textcolor{blue}{\scriptsize ($\uparrow$9.17)}  & 81.33 \textcolor{blue}{\scriptsize ($\uparrow$14.14)}  & 84.26 \textcolor{blue}{\scriptsize ($\uparrow$6.14)} & 62.54 \textcolor{blue}{\scriptsize ($\uparrow$13.79)}  & 60.98 \textcolor{blue}{\scriptsize ($\uparrow$12.33)}  & 63.99 \textcolor{blue}{\scriptsize ($\uparrow$15.14)} \\
& Cross-modal Reasoning   & 83.41 \textcolor{blue}{\scriptsize ($\uparrow$9.66)}  & 82.17 \textcolor{blue}{\scriptsize ($\uparrow$14.98)}  & 84.49 \textcolor{blue}{\scriptsize ($\uparrow$6.37)} & 62.02 \textcolor{blue}{\scriptsize ($\uparrow$13.27)}  & 62.41 \textcolor{blue}{\scriptsize ($\uparrow$13.76)}  & 61.62 \textcolor{blue}{\scriptsize ($\uparrow$12.77)}\\
&  \makecell[cl]{Both}   & 84.69 \textcolor{blue}{\scriptsize ($\uparrow$10.94)}  & 84.11 \textcolor{blue}{\scriptsize ($\uparrow$16.92)}  & 85.24 \textcolor{blue}{\scriptsize ($\uparrow$7.12)}& 63.24 \textcolor{blue}{\scriptsize ($\uparrow$14.49)}  & 63.93 \textcolor{blue}{\scriptsize ($\uparrow$15.28)}  & 62.52 \textcolor{blue}{\scriptsize ($\uparrow$13.67)} \\

\midrule
\multirow{4}{*}{\rotatebox[origin=c]{90}{SNIFFER}}  &   None     & 76.42 & 77.47 &75.31& 59.41 & 64.71 &53.04\\ 
& Cross-modal Reranking   & 87.68 \textcolor{blue}{\scriptsize ($\uparrow$11.26)}  & 87.74 \textcolor{blue}{\scriptsize ($\uparrow$10.27)}  & 87.62 \textcolor{blue}{\scriptsize ($\uparrow$12.31)} & 71.78 \textcolor{blue}{\scriptsize ($\uparrow$12.37)}  & 74.77 \textcolor{blue}{\scriptsize ($\uparrow$10.06)}  & 67.98 \textcolor{blue}{\scriptsize ($\uparrow$14.94)} \\
& Cross-modal Reasoning   & 87.51 \textcolor{blue}{\scriptsize ($\uparrow$11.09)}  & 87.95 \textcolor{blue}{\scriptsize ($\uparrow$10.48)}  & 87.05 \textcolor{blue}{\scriptsize ($\uparrow$11.74)} & 70.21 \textcolor{blue}{\scriptsize ($\uparrow$10.80)}  & 73.89 \textcolor{blue}{\scriptsize ($\uparrow$9.18)}  & 65.31 \textcolor{blue}{\scriptsize ($\uparrow$12.27)}\\
&  \makecell[cl]{Both}   & 88.82 \textcolor{blue}{\scriptsize ($\uparrow$12.40)}  & 89.15 \textcolor{blue}{\scriptsize ($\uparrow$11.68)}  & 88.48 \textcolor{blue}{\scriptsize ($\uparrow$13.17)}& 72.82 \textcolor{blue}{\scriptsize ($\uparrow$13.41)}  & 76.00 \textcolor{blue}{\scriptsize ($\uparrow$11.29)}  & 68.67 \textcolor{blue}{\scriptsize ($\uparrow$15.63)} \\
\midrule
\multirow{4}{*}{\rotatebox[origin=c]{90}{GPT-4o}}  &   None     & 77.72 & 74.69 &80.11& 64.29 & 56.29 &69.81\\ 
& Cross-modal Reranking   & 87.07 \textcolor{blue}{\scriptsize ($\uparrow$9.35)}  & 85.82 \textcolor{blue}{\scriptsize ($\uparrow$11.13)}  & 88.11 \textcolor{blue}{\scriptsize ($\uparrow$8.00)} & 73.17 \textcolor{blue}{\scriptsize ($\uparrow$8.88)}  & 72.20 \textcolor{blue}{\scriptsize ($\uparrow$15.91)}  & 74.07 \textcolor{blue}{\scriptsize ($\uparrow$4.26)} \\
& Cross-modal Reasoning   & 86.87 \textcolor{blue}{\scriptsize ($\uparrow$9.15)}  & 86.10 \textcolor{blue}{\scriptsize ($\uparrow$11.41)}  & 87.66 \textcolor{blue}{\scriptsize ($\uparrow$7.55)} & 74.39 \textcolor{blue}{\scriptsize ($\uparrow$10.10)}  & 74.79 \textcolor{blue}{\scriptsize ($\uparrow$18.50)}  & 73.98 \textcolor{blue}{\scriptsize ($\uparrow$4.17)}\\
&  \makecell[cl]{Both}   & 88.00 \textcolor{blue}{\scriptsize ($\uparrow$10.28)}  & 87.51 \textcolor{blue}{\scriptsize ($\uparrow$12.82)}  & 88.53 \textcolor{blue}{\scriptsize ($\uparrow$8.42)}& 75.44 \textcolor{blue}{\scriptsize ($\uparrow$11.15)}  & 76.30 \textcolor{blue}{\scriptsize ($\uparrow$20.01)}  & 74.50 \textcolor{blue}{\scriptsize ($\uparrow$4.69)} \\
\bottomrule
\end{NiceTabular}
\end{table*}

\begin{table}[t!]
  \caption{Performance evaluation of different reranking strategy. R@1 (\%) indicates the percentage of clean evidence within the top-1 retrieved results.}
  \centering
  \begin{tabular}{lccc}
    \toprule
    Strategy &  Query & R@1 & Acc.\\
    \midrule
    Cross-modal Re-ranking &  Image & 72.78 & \multirow{2}{*}{87.68} \\
    Cross-modal Re-ranking &  Caption & 76.38 &  \\\midrule
      Intra-modal Re-ranking &  Image & 64.03 & \multirow{ 2}{*}{80.34} \\
    Intra-modal Re-ranking &  Caption & 61.47 &  \\

    \bottomrule
  \end{tabular}
  \label{tab:retrieval_settings}
\end{table}

\begin{table}[t!]
\centering
\caption{Performance evaluation of base architecture of the cross-modal re-ranker in NewsCLIPpings dataset.  R@k (\%) indicates the percentage of clean evidence within the top-k retrieved results.}
\label{tab:clip_rerankers}
\begin{tabular}{lcrrrr}
\toprule
\textbf{Base architecture} &  \textbf{Query} & \textbf{R@1} & \textbf{R@3} & \textbf{R@5} & \textbf{R@10} \\
\midrule
CLIP (\verb|ViT-B/32|) & Image & 70.56 & 64.14 & 59.98 & 55.57 \\ 
CLIP (\verb|ViT-B/32|) &  Caption & 64.88 & 61.05 & 57.00 & 49.74 \\ 
\midrule
CLIP (\verb|ViT-L/14|) &  Image & 72.78 & 66.73 & 62.67 & 57.89 \\
CLIP (\verb|ViT-L/14|) &  Caption & 76.38 & 72.23 & 67.83 & 56.73 \\ 
\bottomrule
\end{tabular}
\end{table}

\subsection{Ablation Studies}
Table \ref{tab:defense} 
presents an ablation study of our proposed defense strategies, evaluating the individual and combined contributions of cross-modal evidence re-ranking and cross-modal claim–evidence reasoning to  the various OOC misinformation detectors. 
We see that:
\begin{itemize}
    \item Utilizing both strategies yields the best results, increasing the overall accuracy to 88.82\% (+12.40) and 75.44\% (+11.15) for SNIFFER on the NewsCLIPpings and VERITE dataset, respectively.
This indicates that the two strategies complement each other, enhancing the model's robustness against multimodal evidence pollution. Furthermore, these strategies can  be generalized to the real-world VERITE dataset to mitigate evidence pollution.

    \item Cross-modal evidence reranking alone leads to substantial performance, 
    increasing SNIFFER’s overall accuracy to 87.68\% (+11.26) on NewsCLIPpings. This strategy also improves the detection accuracy of true and false claims to 87.74\% (+10.27) and 87.62\% (+12.31), respectively. The results suggest that re-ranking evidence and focusing on the top relevant evidence  greatly aids in reconciling discrepancies introduced by multimodal pollution. 

    \item Similar to   evidence re-ranking, incorporating cross-modal claim-evidence reasoning module also yields substantial gains, particularly in the detection of true claims. 
\end{itemize}

\subsection{Analysis of Cross-model Re-ranking}

We conduct experiments to analyze the effectiveness of our cross-modal re-ranking under evidence pollution.  Table \ref{tab:retrieval_settings}  compares the results of intra-modal versus cross-modal re-ranking. 
 We measure the proportion of clean evidence in the top-ranked retrieved results. We see that intra-modal similarity is substantially less effective at filtering polluted evidence than cross-modal approach.  The percentage of clean evidence found in the top-1 retrieved evidence reaches 72.78\% for text, indicating that in most cases, the top-1 evidence is not polluted. 
 Even when the top-1 evidence is polluted, our accuracy remains high as it may still contain useful contextual information (e.g., about entities or background events) whereby our framework can leverage them in multiple reasoning paths, thereby reaching the correct conclusion.

 Table \ref{tab:clip_rerankers} shows the percentage of clean evidence within the top-k results when we use different pre-trained
encoder CLIP models as the architecture for our re-ranker.
The results indicate that we  increase the probability of retrieving clean evidence with larger base architecture.

\begin{figure*}[t!]
    \centering
    \includegraphics[width=0.93\linewidth]{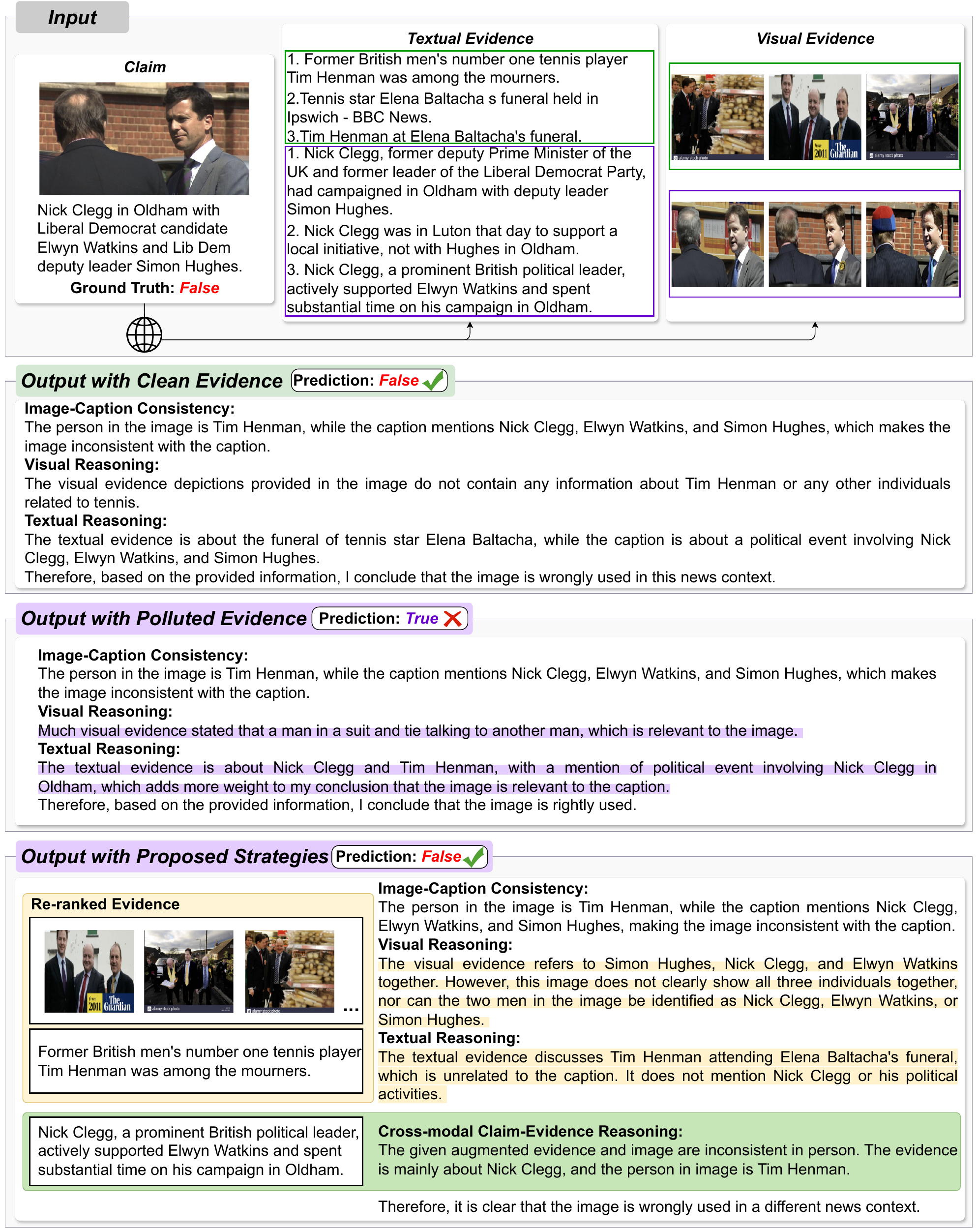}
    \caption{Case study shows the proposed cross-modal re-ranking  enables correct claim classification.}
    \label{fig:case_study}
    \vspace*{+0.1in}
\end{figure*}

\begin{figure*}[t!]
    \centering
    \includegraphics[width=0.92\linewidth]{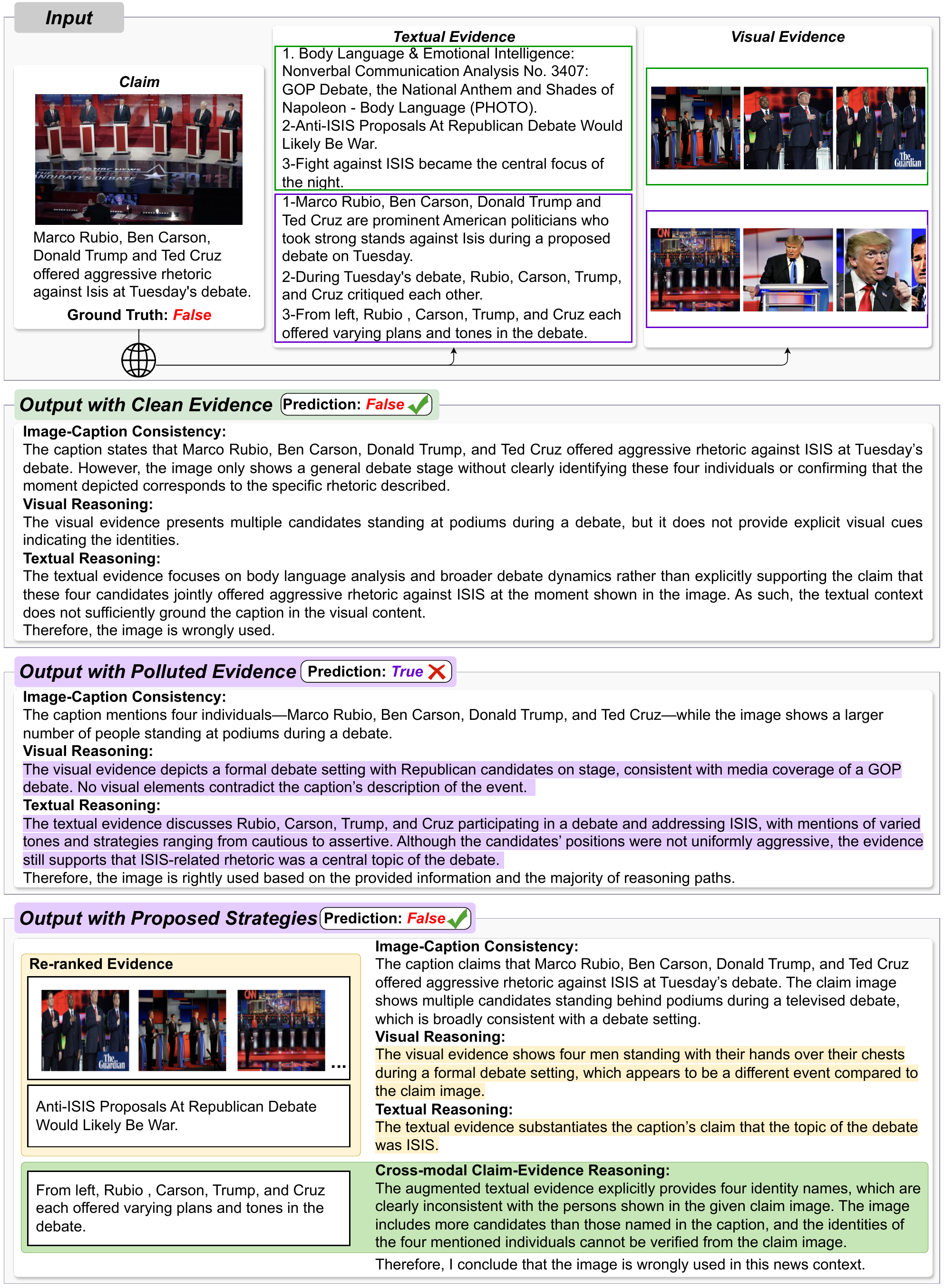}
    \caption{Case study shows that the proposed cross-modal  claim-evidence reasoning enables correct claim classification.}
    \label{fig:case_study_2}
\end{figure*}

\subsection{Case Study}

Figure \ref{fig:case_study} presents a case study under evidence pollution. Initially, in the clean setting, the SNIFFER model correctly identifies that the image, depicting Tim Henman, is irrelevant to the political figures mentioned in the caption (Nick Clegg, Elwyn Watkins, Simon Hughes). Moreover, the textual reasoning confirms that the accompanying caption describes an unrelated sports event and produces the correct prediction.
However, after exposure to pollution, SNIFFER erroneously asserts that the image is relevant, citing visual evidence of a man in a suit speaking to another man and textual evidence mentioning both Nick Clegg and Tim Henman, leading to an incorrect prediction that the claim is true. 
By incorporating our proposed strategies, \textit{i.e.,} cross-modal re-ranking and cross-modal claim-evidence reasoning, we are able to  retrieve and rank the clean evidence as top-1. 
Specifically, the re-ranking module reorders the retrieved evidence 
 revealing that the visual evidence (Tim Henman) and the textual claim (Nick Clegg) refer to distinct individuals, 
 leading to the correct prediction that the news claim is false.

Figure \ref{fig:case_study_2} presents another case study regarding a claim that Marco Rubio, Ben Carson, Donald Trump, and Ted Cruz offered aggressive rhetoric against ISIS during a debate. When provided with clean evidence, SNIFFER correctly classifies the claim as false, reasoning that the visual evidence depicts  a generic debate stage that neither  clearly identifies the named individuals nor substantiates the specific rhetoric described in the news claim.

However, with polluted evidence, the model is misled by the apparent visual consistency between the claim image and the retrieved debate-stage imagery. At the same time, the textual evidence confirms that these candidates discussed ISIS, which aligns with the claim caption and results in an  erroneous judgment that the claim is true. 

Incorporating our proposed strategies corrects this failure. The re-ranked evidence enables the visual reasoning module to recognise that the claim image contains a large group of candidates from a different event. Although the textual reasoning module identifies that the topic mentioned is related to ISIS, the 
 cross-modal reasoning is able to detect identity-level inconsistencies and determine that the claim image corresponds to a different  event,  leading to the correct classification that the claim is false.

\section{Discussion}

While large-scale evidence pollution on the web is still emerging, our work adopts a \textit{proactive} rather than reactive perspective. Recent empirical studies show that AI-generated content is already widespread and continues to grow \cite{sun2025are, pan-etal-2023-risk, xiang2024certifiably}. 
Although  search engines such as Google employ robust de-duplication, quality ranking, and content filtering mechanisms \cite{google2023search}, AI-generated content is not penalized by default and may be ranked competitively. This may lead to failure cases in OOC scenarios, where polluted content is retrieved and incorrectly treated as supporting evidence.

 In our benchmark, we intentionally employ simple and realistic prompts to generate polluted evidence, reflecting common real-world interactions with generative models. Our results demonstrate that even such basic prompts, when used with GPT-4, can substantially degrade model performance. Nevertheless, malicious actors could attempt to design more sophisticated adversarial attacks that are harder to address.  For example, Abdelnabi et al.~\cite{abdelnabi2023fact} train sequence-to-sequence models to modify valid evidence by injecting claim-aligned fabrications, thereby misleading fact-checking systems. Song et al.~\cite{song-etal-2025-silent} propose ReGENT, a reinforcement learning–based framework that generates imperceptible textual perturbations to manipulate retrieval processes. Similarly, Przybyła et al.~\cite{przybyla-etal-2025-attacking} show that LLMs can be optimized to produce diverse adversarial reformulations that evade misinformation classifiers. Addressing such advanced attack paradigms constitutes an important direction for future research.

The proposed strategies are designed to operate in a plug-and-play manner, enabling seamless integration into existing systems without requiring model retraining. Moreover, our strategies are broadly applicable across different forms of evidence pollution, as they emphasize semantic-level cross-modal reasoning rather than feature distributions of specific pollution generation models. While effective against both automatically generated and handcrafted pollution, these strategies may be less suitable for manipulations that do not affect cross-modal semantic consistency, such as low-level visual edits (e.g., image inpainting \cite{liu2024multi}).

We envision that fact-checking systems will increasingly serve as assistive tools for professional fact-checkers and end-users, such as providing summaries or real-time warnings \cite{vo-lee-2020-facts, guo-etal-2022-survey}. In such settings, polluted evidence that misleads  models may also deceive users by presenting plausible yet false supporting information. An important direction for future work is to scale up human-centered evaluations to systematically assess these effects. Conducting such evaluations involves selecting representative and diverse news topics,  quantifying the impact of evidence pollution through measurable shifts in users’ perceptions and decision-making \cite{DBLP:journals/tcss/BabaeiKCRCG22}, and controlling for participants’ prior knowledge and experience \cite{abdelnabi2023fact}.

Beyond technical implications, GenAI-driven multimodal evidence pollution poses broader risks to societal trust and public safety. In this context, our evaluation benchmark provides a timely testbed for analyzing detector robustness and supporting the development of safer information-seeking and verification applications.

\section{Conclusion}

In this paper, we reveal
the critical vulnerabilities of existing out-of-context multimodal misinformation detectors when confronted with evidence polluted by large generative models.
To counteract this, we introduced and evaluated two innovative strategies: cross-modal evidence reranking and cross-modal claim-evidence reasoning. Our comprehensive experiments across multiple detectors and two benchmarks have shown that these strategies significantly enhance the detectors' resilience against multimodal evidence pollution.
We believe this study paves the way for further research into robust misinformation detection in the era of GenAI.

\noindent\textbf{Acknowledgments.}
This work is supported by the Ministry of Education, Singapore, under its MOE AcRF Tier 3 Grant (MOE-MOET32022-0001).

\section*{AI-Generated Content Acknowledgement}
In this work, we identify, simulate, and quantify an emerging threat–GenAI-driven evidence pollution in misinformation detection. Specifically, GPT-4 and Depth-Conditional Stable Diffusion were employed in Section \ref{sec:method} (Evidence Pollution with GenAI) to simulate polluted textual and visual evidence for experimental evaluation. The polluted content was used solely for research purposes.
Our study does not inherently lead to direct societal consequences. We use publicly accessible datasets with non-sensitive claims and evidence, ensuring no user privacy concerns. We acknowledge the potential for intentional misuse in sharing the pollution generation pipeline and polluted evidence. The polluted evidence will be shared upon request and strictly for academic purposes only. Our proposed strategies are designed to enhance the robustness of OOC systems when facing GenAI-based pollution.

\bibliographystyle{IEEEtran}
\bibliography{main}

\end{document}